\documentclass{article}

\usepackage{arxiv}

\usepackage[utf8]{inputenc} 
\usepackage[T1]{fontenc}    
\usepackage{hyperref}       
\usepackage{url}            
\usepackage{booktabs}       
\usepackage{amsfonts}       
\usepackage{nicefrac}       
\usepackage{microtype}      
\usepackage{lipsum}
\usepackage[bottom]{footmisc}
\usepackage{amsmath,amssymb,amsfonts}
\usepackage{algorithmic}
\usepackage{url}
\usepackage{graphicx}
\usepackage{multirow}
\usepackage[document]{ragged2e}
\usepackage{makecell}
\title{Probabilistic modelling of gait \\
 for robust passive monitoring in daily life\\}

\author{\large{Yordan P. Raykov$^1$, Luc J.W. Evers$^{2,3}$, Reham Badawy$^4$, Bastiaan
Bloem$^2$,}\\
\large{\textbf{Tom M. Heskes$^3$, Marjan Meinders$^5$, Kasper Claes$^6$, Max A. Little$^{4,7}$}}}

\begin{document}
\maketitle
\begin{abstract}
Passive monitoring in daily life may provide invaluable insights about
a person's health throughout the day. Wearable sensor devices are
likely to play a key role in enabling such monitoring in a non-obtrusive
fashion. However, sensor data collected in daily life reflects
multiple health and behavior-related factors together.
This creates the need for structured principled analysis to produce
reliable and interpretable predictions that can be used to support clinical
diagnosis and treatment. In this work we develop a principled modelling
approach for free-living gait (walking) analysis. Gait is a promising target
for non-obtrusive monitoring because it is common and indicative of various
movement disorders such as Parkinson's disease (PD), yet its analysis
has largely been limited to experimentally controlled lab settings. To locate and
characterize stationary gait segments in free living
using accelerometers, we present an unsupervised statistical framework
designed to segment signals into differing gait and non-gait
patterns. Our flexible probabilistic framework combines empirical assumptions about gait into a principled
graphical model with all of its merits. We demonstrate the approach
on a new video-referenced dataset including unscripted daily living
activities of 25 PD patients and 25 controls, in and around their own houses.
We evaluate our ability to detect gait and predict medication induced
fluctuations in PD patients based on modelled gait. Our evaluation
includes a comparison between sensors attached at multiple body locations including wrist,
ankle, trouser pocket and lower back.
\end{abstract}
1 - Aston University; School of Engineering and Applied Sciences; Department of Mathematics, Birmingham, United Kingdom\\
2 - Radboud university medical center; Donders Institute for Brain, Cognition and Behaviour; Department of Neurology; Center of Expertise for Parkinson $\&$ Movement Disorders; Nijmegen, The Netherlands\\
3 - Radboud University; Institute for Computing and Information Sciences, Nijmegen, The Netherlands\\
4 - University of Birmingham; School of Computer Science, Birmingham, United Kingdom\\
5 - Radboud university medical center; Radboud Institute for Health Sciences; Scientific Center for Quality of Healthcare (IQ healthcare), Nijmegen, The Netherlands\\
6 - UCB Biopharma, Brussels, Belgium\\
7 - Media Lab, Massachusetts Institute of Technology, Cambridge, Massachusetts, United States of America\\

\section{Introduction}
Ubiquitous consumer devices such as smartphones and wearables are
equipped with low power inertial sensors such as accelerometers and gyroscopes
capable of continuously recording their wearer's movements. In controlled
laboratory settings, such sensors have been used successfully to measure
symptoms of patients with various movement disorders, such as Parkinson's
disease (PD) \cite{Lipsmeier2017,sha2008spa,horak2013objective}.
However, these measurements only provide a snapshot of the patient's
condition, and may not be representative of the symptoms experienced
in daily living conditions outside the lab, for example because of observer effects \cite{warmerdam2020long}.
Unobtrusive wearable sensors enable us to monitor patients in daily
life, which may provide patients, care providers and researchers with
useful insights in the course of symptoms \cite{De2016}.

However, obtaining reliable and interpretable measurements in uncontrolled
environments is difficult. One strategy has been to record the patient's
ability to perform specific tasks (e.g. walk 10 meters) at different
times of the day (\textit{active tests}) \cite{arora2014high}. However,
an important limitation of active tests is that patients
are interrupted during the tests, which can lead to high attrition
in compliance \cite{bot2016mpower}. Additionally, it is practically
impossible to obtain a continuous view of symptom fluctuations using
short active tests.

Instead of instructing patients to perform specific tasks, we could
use daily routine activities that are affected by the patient's condition
to measure how someone's symptoms fluctuate throughout the day (i.e \textit{
passive monitoring}). An important example of such activity is walking,
otherwise known as \emph{gait}. Many movement disorders are associated
with alterations in gait patterns, and neurologists often use in-clinic
gait examination to establish a diagnosis. PD-related changes in 
gait patterns consist of continuous impairments involving slowness
and reduced arm swing (\textit{bradykinetic gait}) and episodic hesitations
to produce effective steps (\textit{freezing of gait}). In many patients,
bradykinetic gait is already present early in the disease \cite{del2019gait}
and is responsive to symptomatic medication (e.g. levodopa) \cite{curtze2015levodopa}.
Therefore, measuring free living gait could serve as a marker for
disease progression and therapy-related symptom fluctuations in PD
patients. This would allow for unobtrusive remote patient monitoring,
and can potentially facilitate titration of medication, early diagnosis
and evaluation of new drugs \cite{Hodgins2008}.

In order to extract meaningful information about a patient's free
living gait, we need a robust framework for gait detection and characterization
of the gait pattern. When it comes to gait detection, most
existing work can be classified into two systems: (1) \emph{activity
recognition} systems which classify (real time) data into a fixed
number of pre-specified activities \cite{Kwapisz2011activity,Bao2004activity,Ravi2005activity},
and (2) \emph{gait detection} systems that perform binary classification
to determine whether a window of data belongs to the gait or no gait
class\cite{hanlon2009real,brajdic2013,Williamson2000gait}. Systems
in both categories are typically trained and evaluated using labelled
data from a pre-defined, and sufficiently distinguishable set of scripted
activities, often collected in controlled environments. However, in uncontrolled
environments, it is practically impossible to anticipate all the activities
users might engage in. This means we expect a distributional mismatch between the activity
data available during training and the actual out-of-sample data. In particular when such systems 
use many different features or deep learning, it is difficult to anticipate how the algorithm will cope 
with unseen activities. Furthermore, in health
monitoring applications, we expect large differences in the
way patients perform the same activity, gait in particular, due to their disease symptoms.
This means that, on the one hand, we need labelled training data that better reflects real-life variation and variation due to disease symptoms. On the other hand, we need to acknowledge that it will remain infeasible to capture all variation in training data sets, and we need models that can account for this.  

Binary gait detection systems are often implemented using threshold values applied to statistical summaries
of windowed data\cite {Sama2012,Tao2012,brajdic2013,Kavanagh2008,del2016}. Whereas this
``low complexity'' approach may have acceptable accuracy to globally describe how much users walk,
problems can emerge when it is used as a starting point for evaluating the quality of the gait in health monitoring applications.
For example, these systems group all gait together, regardless of changes in the gait pattern that can occur even within the same gait segment (e.g. because of changes in symptoms, pace or environment). As a
result, the detected gait segments are likely heterogeneous and non-stationary,
introducing unpredictable biases into subsequent gait pattern analysis. 

In this work, we propose a unified framework for gait detection and gait pattern
analysis. We have combined some the most common criteria used for
gait detection into a principled probabilistic graphical model, which
can be directly applied to the accelerometer data to infer varying gait and
non-gait patterns occurring in free living. We adopt a flexible \textit{nonparametric}
model which can locate different gait and non-gait activities that vary
both in terms of their statistical and temporal characteristics. Specifically,
we use a set of high order \textit{autoregressive}
(AR) \textit{processes}. The AR process is a parametric model of the frequency
spectrum, hence it directly captures characteristics derived from the \textit{power
spectral density} of the data. At the same time, AR processes are
time domain models which allows us to couple them with a \textit{nonparametric}
hidden Markov model (HMM) leading to an AR-iHMM also known as a \textit{nonparametric switching AR process} \cite{fox2009nonparametric}) to capture the longer-term changes in behavior patterns and gait types in free living conditions.

\begin{figure*}[htbp]  
\center
\includegraphics[width=0.995\columnwidth]{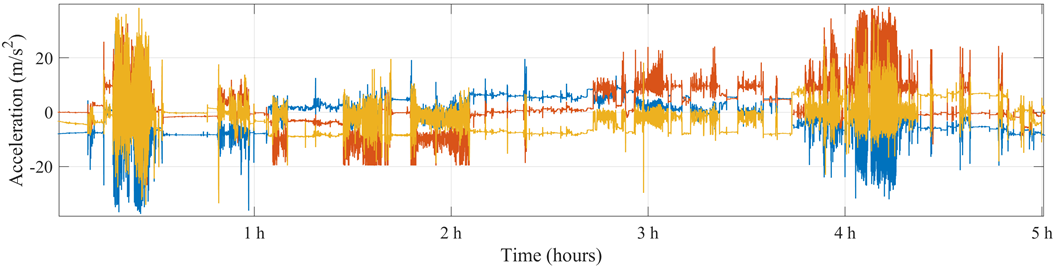}

\caption{Example of 3-axis accelerometer data collected during unscripted activities in and around the home using a smartphone placed in a PD patient's front trouser pocket.}
\label{fig.AccelerometerData} 
\end{figure*}

Different HMMs have been applied previously, both for generic
activity recognition \cite{lee2011activity, lukowicz2004recognizing, olguin2006human, Mannini2011activity, Mannini2013},
binary gait classification \cite{nickel2011benchmarking} and sub-typing
of gait \cite{Mannini2011gait}. However, this has been done in a
supervised setting where a parametric HMM is trained on features extracted
from the windowed sensor data. By contrast, the AR-iHMM proposed here
can be directly applied to the pre-processed time series, which allows us
to circumvent challenges related to spectral estimation of windowed signals, and avoid the
need for engineering many features.

Once free living data is segmented into different states, they can be used to separate gait from non-gait data, but also to identify segments in time of sufficiently different gait as well as short-term events
causing interruption of gait. To demonstrate the applicability of this analytical framework, we
use a new, unique dataset consisting of sensor data from various wearables
and concurrent reference video annotations, collected during unscripted
daily living activities in and around the homes of 25 patients with
PD, and 25 age-matched controls. Our results demonstrate state-of-the-art
accuracy at detecting healthy and pathological gait in free living conditions
from different sensor wear locations. Furthermore, we show that the model
can identify changes in gait pattern after medication intake in
individuals with PD.

\section{Related work\label{sec:Related-work}}

In the last two decades, advances in wearable sensors have made it
feasible to unobtrusively monitor patients outside controlled laboratory
conditions, allowing us to study real-life gait patterns. However,
to successfully deliver on that promise, we need tools which can reliably
and robustly model data recorded from wearables in this setting. Here
we review relevant prior work in terms of device location, gait detection
algorithms, and gait characteristics under study.

\paragraph{Device location}

Studies on gait detection and gait pattern analysis have used accelerometers,
gyroscopes and/or magnetometers worn on various body locations, including
the trouser pocket \cite{brajdic2013}, the lower back \cite{Wong2007},
the shin or ankle \cite{Mannini2013}, the shoe \cite{klucken2013unbiased},
as well as the wrist \cite{Doherty2017}. The choice of device location
is influenced by the expected gait detection accuracy, the type of
gait characteristics that can be reliably estimated, patient acceptance,
and the commercial availability of devices. An extensive review of
widely-used wearable devices and their sensors for gait analysis can
be found in Tao et al. \cite{Tao2012}, and a focused review on sensor
placement for monitoring of PD can be found in Brognara et al. \cite{brognara2019assessing}.

There is no consensus on the best device location to detect and
characterize the gait of PD patients, and whether there is added value
in combining multiple locations. Therefore, we evaluate our proposed framework on various commonly
used sensor locations.

Another concern can be the limited commercial availability and high
costs of ``research-grade'' devices. For this reason, we include a consumer
smartphone in our comparison, which is widely available
and relatively low-cost.

\paragraph{Gait detection}

Most gait detection techniques rely on parametric assumptions about
the spectral density, time domain distribution or
both \cite{brajdic2013}. Typically, features are extracted from
windows of fixed width, and the decision to classify a window as gait
or non-gait behavior is made using pre-defined thresholds or using
a trained classifier. For example, one of the most widely used methods
for identifying gait estimates the standard deviation of a windowed
accelerometer signal, and uses a fixed threshold value \cite{Randell2003}.
An alternative, and similarly popular approach is the window-based
analysis of spectral features \cite{Kavanagh2008,Tao2012,Dijkstra2010detection}.
Gait is typically highly periodic with Nyquist bandwidth of 10-15Hz
\cite{Antonsson1985}. This has motivated the use of the \textit{short-time
Fourier transform} (STFT) \cite{Allen1977} to detect gait. For example,
Sama et al. \cite{Sama2017, Sama2012} studied the energy of the
accelerometer signal in 800 different frequency bands. They applied
Relief feature selection to identify the energy bands that are
most descriptive of gait and then they used a support vector machine
(SVM) to detect gait. Karantonis et al. \cite{Karantonis2006} suggested directly analyzing
the Fourier coefficients of the z-axis on the accelerometer to look
for \textit{sufficient} power at the expected range of walking frequencies
(0.7\textendash 3.0 Hz). The time-frequency resolution issues of STFT-based
walking detection have sometimes been addressed using wavelet transforms
\cite{mallat2008wavelet}. Continuous wavelet transforms often require
large computational effort \cite{Figo2010}, but discrete wavelet
transforms can be used to efficiently estimate high quality features
of gait \cite{El2018}, more efficiently even compared to Fourier
transform \cite[page 254]{little2019machine}. We can also encode
the power spectrum directly in the time domain if we use windowed auto-correlation
\cite{Rai2012} and then use the values at a subset of time lags
corresponding to the duration of the gait cycle \cite{Makihara2010, Rai2012}.

A problem with these different window-based feature extraction methods
is that signals acquired in daily life are highly non-stationary.
When these non-stationarities occur within a window, for example, the
transition from standing to gait, they may reduce the usefulness of the
extracted features, particularly in the case of STFT (as we will further
discuss in Section \ref{sec:Challenges of modelling gait}).

These different gait detection systems not only vary in the features
they rely on, but also in the classification algorithm they use. ``Traditional''
classification techniques such as support vector machines and random
forest classifiers are commonly trained on window-based features \cite{nyan2006classification,mathie2004classification}.
In addition, HMMs have also been used to detect
gait based on window features, which offers the potential advantage
of incorporating the sequential nature of human behavior \cite{nickel2011using,Mannini2011activity,haji2018segmentation}. 
Haji et al. \cite{haji2018segmentation}
trained a hierarchical HMM on frame variance, raw data and second
order polynomial coefficients and demonstrated that, in more challenging
settings, this method significantly improves gait detection compared
to, for example, peak detection and dynamic time warping (explained
below). Despite the heterogeneity in gait patterns, gait detection
is generally treated as a binary classification problem (gait/non-gait).

Other approaches avoid feature extraction altogether. For example,
a \textit{stride template} can be formed offline and online similarity
to the template be determined (e.g. via cross-correlation \cite{Ying2007,Marschollek2008}
or dynamic time warping \cite{Rong2007}). However, this approach
is not very practical for detection of pathological gait whose temporal
pattern can vary significantly even within the same individual. More
recently, generic activity recognition pipelines based on deep learning
methods \cite{camps2018deep,hammerla2016} are being introduced,
although the scarcity of labelled free living data currently limits
their practical use.

\paragraph{Characterization of the gait pattern}

Once gait episodes have been identified, studies have used various
approaches to characterize the gait pattern in movement
disorders such as PD. Many studies try to identify important events
of the gait cycle, including the heel strike or initial contact (IC),
and final contact (FC) of both feet. Several variations to peak detection
have been used for this, which may benefit from pre-processing the
acceleration signal using continuous wavelet transforms (CWT) \cite{Mccamley2012enhanced}. The timing of IC and FC events is then used to compute temporal gait features
such as step time, swing time, stance time, and double support time.
Additionally, based on assumptions about the exact sensor positioning
and the biomechanics of gait, location-specific algorithms can be
used to estimate spatial gait features. For example, having identified
the ICs and FCs, one can use the inverted pendulum model to estimate
the step length from the accelerometer signal of a sensor on the lower
back \cite{zijlstra2003assessment}. Del Din et al. \cite{del2015validation} used
this approach and showed that free living gait analysis discriminated
better between PD patients and healthy controls than lab-based gait
analysis, which illustrates the potential of free living gait analysis.
Moore et al. \cite{Moore2011} suggested that the step length estimated using
an ankle sensor could be used to track the free living gait pattern
of PD patients, but only included three PD patients monitored over
24 hours in an apartment-like setting.

Other approaches focus on analyzing the periodicity of the accelerometer
signal during gait, either based on the PSD or auto-correlation in
the time domain. An advantage of these methods is that they are less
dependent on location-specific assumptions, compared to identifying
gait cycle events and computing the step length. For example, Weiss et al. \cite{weiss2014objective}
computed the width of the dominant frequency in the PSD during free
living gait (based on the accelerometer signal from a lower back sensor),
and demonstrated that it could be used to predict future falls in
patients with PD. Similarly, Rispens et al. \cite{rispens2015identification} computed
the PSD during free living gait based on a lower back accelerometer,
and showed that the spectral power in the lower frequencies, and the
amplitude and slope of the dominant frequency, were related to the
number of falls in older adults. P{\'e}rez-L{\'o}pez et al. \cite{perez2016assessing} combined
the identification of ICs with analysis of the PSD during individual
strides, and showed that the power in the gait range (based on a waist
accelerometer) was correlated to changes after medication intake in
PD patients. Bellanca et al. \cite{bellanca2013harmonic} suggested that the harmonic
ratio (ratio of the sum of the amplitudes of the even and uneven harmonics,
computing over the PSD of a single stride) could be used as a measure
of step symmetry. Alternatively, the periodicity of free living gait
can also be analyzed in the time domain, for example by estimating
the auto-correlation \cite{moe2004estimation}. All analyses mentioned
in this paragraph strongly depend on accurate localization of stationary
gait segments, which may be sub-optimal given current gait
detection algorithms. In this work, we propose that free living gait
analysis can be improved by employing a unified approach to gait detection
and gait pattern characterization.

\section{Free living data collection\label{sec:Data description}}

Most gait modelling pipelines are both designed and tested on data
recorded under controlled lab conditions. In order to allow for a
more realistic understanding of the challenges of modelling free living
gait data, we have used a new reference dataset from the Parkinson@Home
validation study. This study includes sensor data and video recordings
during uninterrupted and unscripted daily life activities in the participants'
natural environment. In brief, both patients with Parkinson\textquoteright s
disease (PD group) and 25 age-matched participants without PD (non-PD
group) were recruited. Inclusion criteria for both groups consisted
of: (1) age 30 years or older and (2) in possession of a smartphone
running on Android OS version 4.4 or higher. Additional inclusion
criteria for participants in the PD group were: (1) diagnosed with PD by a neurologist, 
(2) receiving treatment with dopaminergic medication
(levodopa and/or dopamine agonist), (3) experiencing motor fluctuations
(MDS-UPDRS item 4.3 \ensuremath{\ge} 1), and (4) known to have PD-related
gait abnormalities, i.e. bradykinetic and/or freezing of gait (MDS-UPDRS
item 2.12 \ensuremath{\ge} 1 and/or item 2.13 \ensuremath{\ge} 1).
PD patients who received advanced treatment (deep brain stimulation
and/or intestinal infusion of levodopa or apomorphine) were excluded.
\begin{figure*}[t]  
\center
\includegraphics[width=0.75\columnwidth]{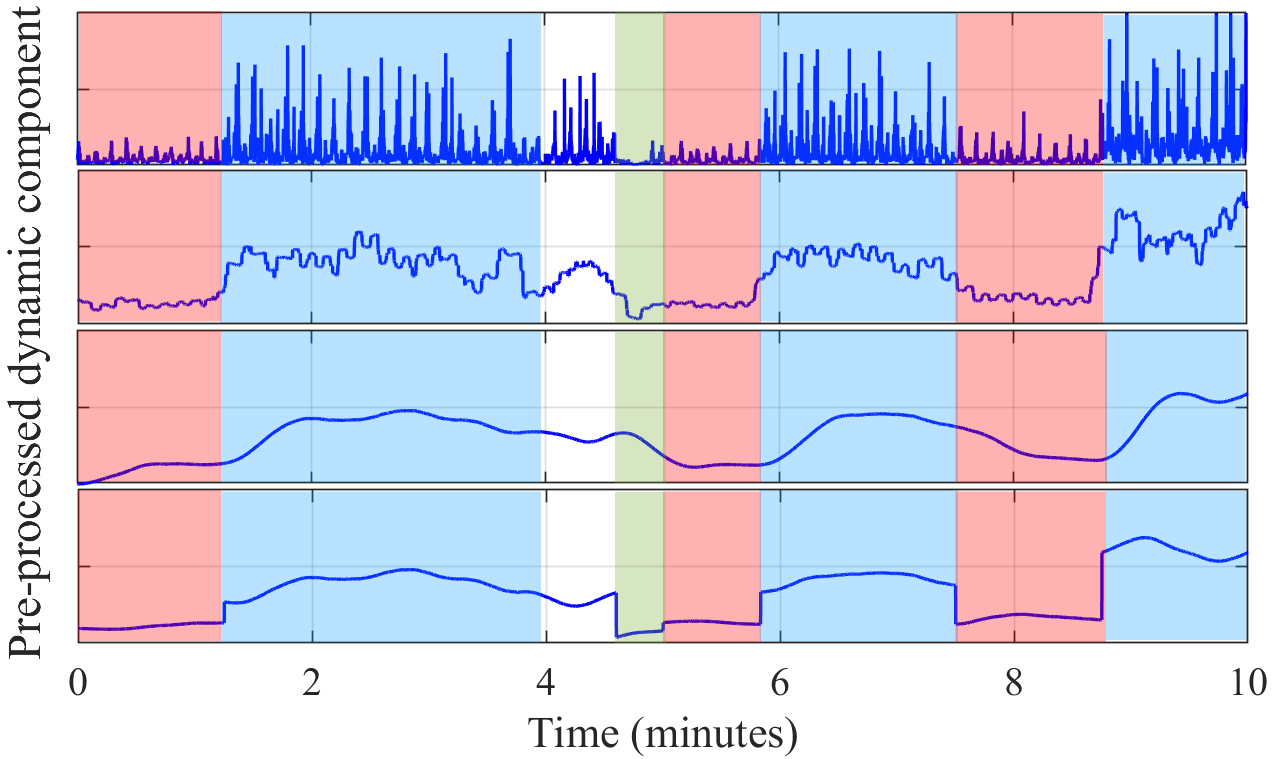}

\caption{Illustrative example of feature smoothing over accelerometer data during unscripted gait. The top panel displays the magnitude of 10 minutes of pre-processed accelerometer data collected using a smartphone placed in the front trouser pocket. The next panel below displays the standard deviation over 1 second windows, which is a descriptive feature of the switching gait patterns in the considered example. In the following panels we display the filtered values of the feature using: standard moving average in time (5 seconds); ``conditional'' moving average with same window, conditioned on the states identified by the AR-iHMM (displayed in red, blue and green). Note: the white segment contains multiple rapidly switching states as identified by the AR-iHMM (not shown).} 
\label{fig.GaitGroupPSD} 
\end{figure*}

Participants were visited in their own homes and each visit included
a standardized clinical assessment (full MDS-UPDRS \cite{Goetz2008}
and AIMS \cite{munetz1988examine}) and an unscripted free living
assessment of at least one hour. To ensure indicative behaviors such
as longer gait cycles were captured, assessors encouraged participants
to include these in their routines. Participants in the PD group were
asked to skip their morning dose of dopaminergic medication before
the visit, so that they were in the OFF medication state at the start
of the visit. After the MDS-UPDRS part III (motor examination) and
free living assessment were conducted in the OFF state, participants
took their usual medication and the full MDS-UPDRS, AIMS and free
living assessment were performed in the ON state, i.e. with the symptomatic
effects of medication present.

During the full visit, participants wore various light-weight sensors
on different body locations. In this study, we used the accelerometer
data from the smartphone worn in the front trouser hip pocket (collected
using the HopkinsPD app \cite{zhan2016high}; all participants were instructed 
to wear trousers with a front pocket), and the accelerometer data from Physilog 4 devices worn on
both ankles, both wrists and the lower back. To allow for time synchronization,
all devices were triggered together (hit ten times against a table)
in front of the video camera at the beginning and end of data collection.

The video recordings during the free living assessments were annotated
by a research assistant, who labeled as ``gait'' any activity that involved at
least 5 consecutive steps, with the exception of any running
episodes.

\section{Challenges of modelling free living gait\label{sec:Challenges of modelling gait}}

Because of its simplicity, robustness and affordability, the 3-axis
accelerometer is by far the most widely used sensor for free living
gait analysis. The accelerometer sensor measures the vector sum of
all sources of acceleration acting on the device in each spatial direction.
The unit of measurement is $m/s^{2}$ and if the device is not under
other sources of acceleration, the only acceleration measured by the
device is due to the force of gravity (zero magnitude under free-fall).
An example of accelerometer data collected during the free living assessment
is shown in Figure \ref{fig.AccelerometerData}.

Analysis of free-living gait is challenging because accelerometer data simultaneously reflects both disease symptoms, behaviour, device orientation, sensor location and environment. This makes it difficult to design a reliable analytical pipeline which untangles these factors and allows us to focus solely on representative aspects of the gait that are relevant for monitoring PD. To make a necessary step in this direction, we highlight some of the common \textit{estimation} challenges which we aim to address. We use examples from the unscripted free living assessments of the Parkinson@Home validation study.

\paragraph{Device orientation}

Accelerometers measure any forces due to accelerations
which partly prevent the device from free-fall in the Earth\textquoteright s
gravitational field. If we are interested
in monitoring gait, however, we first need to remove this field effect
from the raw accelerometer data, as irrelevant device rotations (e.g.
slight variations in the attachment of the sensor) may otherwise confound
any inferences we make about a person's gait. This analytical step
is most commonly done using fusion of data from a magnetometer, gyroscope
and accelerometer \cite{Madgwick2011}, or simply using a digital
low pass filter \cite{Van2013} applied to the accelerometer signal.
Sensor fusion is well justified from a physical modelling point of
view, but it is not very accurate in practice during fast motion \cite{bergamini2014estimating}.
On the other hand, low pass filters are poorly justified, since unwanted
orientation changes can be rapid with a broad bandwidth leading to
unwanted distortions in the time domain depending on the cut off frequency
of the filter. In this work we opt for a piecewise $l_{1}$- trend
filter as motivated in Badawy et al. \cite{badawy2018} which assumes that changes
due to orientation are piecewise linear \cite{kim2009ell_1}.

The accelerometer data we use in any subsequent analysis, is pre-processed
by interpolating to a uniform sample rate (i.e. using cubic spline
interpolation \cite{mckinley1998cubic})), applying the $l_{1}$-
trend filter to each individual axis and computing the magnitude of
acceleration according to $\sqrt{a_{x}^{2}+a_{y}^{2}+a_{z}^{2}}$.

\begin{figure}[htbp]  

\includegraphics[width=0.495\columnwidth]{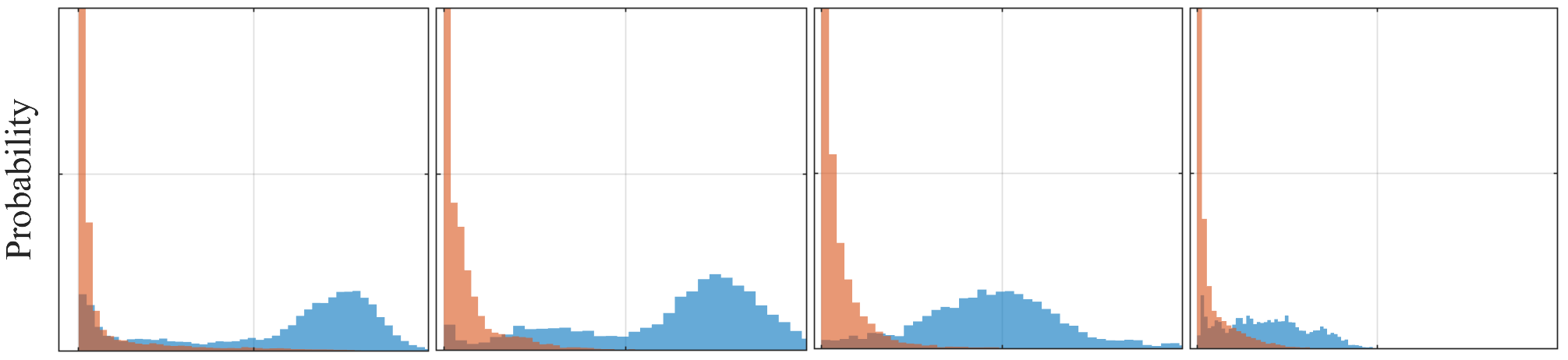}
\includegraphics[width=0.498\columnwidth]{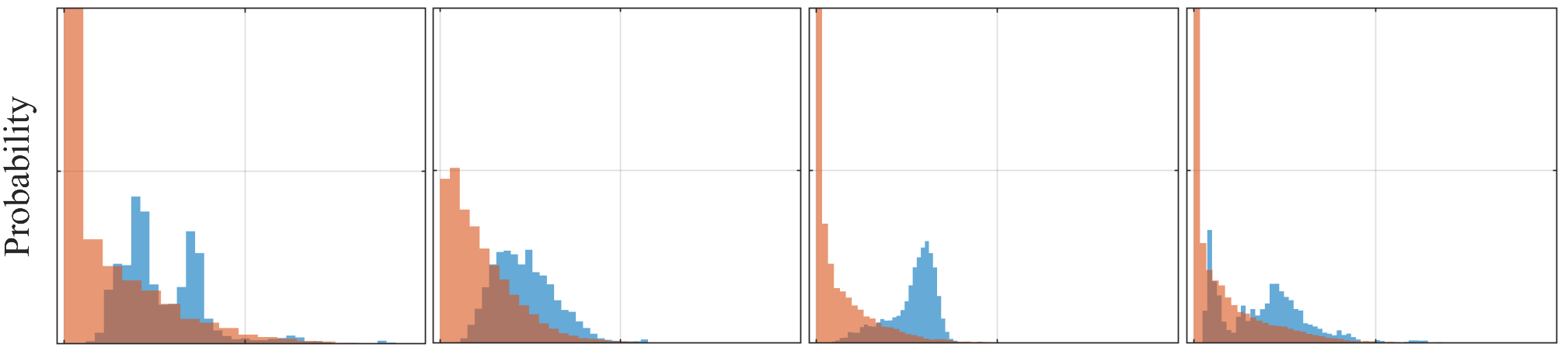}

\caption{Histograms of 1 second window standard deviation of the magnitude of acceleration from different PD patients during the unscripted free living assessment, collected using a smartphone placed in the front trouser pocket (top row) and a wrist-worn wearable device (bottom row). The horizontal axes show the window standard deviation, and the vertical axes show the normalized bin counts. Blue: gait class (video annotations). Orange: non-gait class (video annotations).}
\label{fig.WindowSTDsmartphone} 
\end{figure}

\paragraph{Parsimonious representation of gait data\label{par:Gait-data-distribution}}

We have seen in Section \ref{sec:Related-work} that most pipelines for analysis of gait data involve windowing of the sensor data and estimation of some statistical or spectral features. The estimated feature values are then used to make inferences about the behaviour monitored at that point in time (i.e. gait vs non-gait) or the gait pattern. However, in free living the variability of these features is large. For example, in Figure \ref{fig.WindowSTDsmartphone} we show how much the window standard deviation varies for both gait and non-gait classes, even within a single individual. To reduce some of this variability, we tend to aggregate feature values (i.e. across time, across individuals, across similar behaviours). The way we make such aggregation will inevitably affect the quality of the inferences we make. 

Let us consider the following example: we have $10$ minutes of consecutive gait data from a PD patient where the gait varies significantly across different segments. In Figure \ref{fig.GaitGroupPSD} We plot how much for example the (1 second) window standard deviation varies in time and how the variation can be reduced by smoothing through time (using moving average). The underlying assumption is that feature values collected closely in time, should be similar (i.e. change smoothly). However, when monitoring heterogeneous behaviours such as gait, this assumption does not hold. In contrast, we also display the ``conditional'' moving average obtained if we first segment the series into 4 stationary states (using the proposed AR-iHMM) and only smooth data that belong to the same state.

\begin{figure}[t]  
\center
\includegraphics[width=0.5\columnwidth]{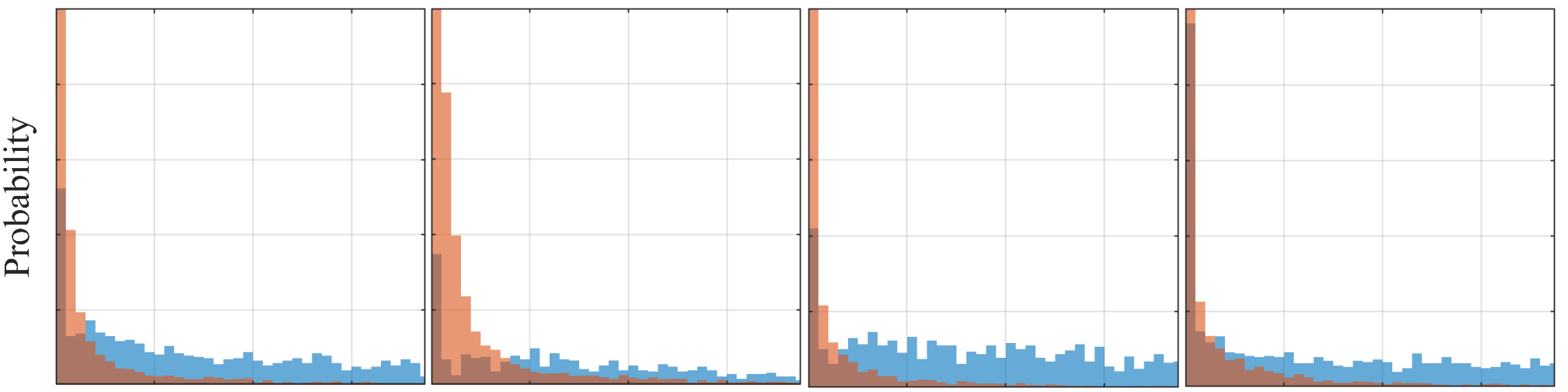}

\caption{Histograms of the spectral energy at typical gait frequencies (0.5 - 10Hz) obtained using STFT with window length 1 second. Each subplot displays the feature distribution for the unscripted free living assessment of a single PD patient, from a smartphone worn in the front trouser pocket. The horizontal axes are different total energy values, and the vertical axes show the normalized bin counts. Blue: gait class, orange: non-gait (according to video annotations).}
\label{fig.WindowMaxFourierCoeff} 
\end{figure}

A similar argument can be made for the features based on the spectrum. A common ``building block'' for the estimation of the spectrum is the \textit{short-time Fourier transform} (STFT). Gait detection based on STFT assumes that if the data contains sufficient spectral power within the range of normal gait cadence, it represents gait activity \cite{brajdic2013}. However, the support at the different frequencies also exhibits large variation both within gait and non-gait classes, as shown in Figure \ref{fig.WindowMaxFourierCoeff}. Similar to standard deviation, more stable spectral estimation can be done by aggregating across neighbouring windows, e.g.
using Welch's overlapped averaging power spectral density (PSD) estimator
\cite[Section 7.4]{Welch1967,little2019machine}. The problem is that this still assumes that the signal is stationary across the windows, which is often not the case in free living data because of abrupt changes in behaviour (e.g. changing pace, turning, starting to make gestures), environment (e.g. changing walking terrain), and PD symptoms (e.g. hesitations to walk through doorways) which affect the characteristics of the gait pattern. If we ignore such variability when choosing the gait segment boundaries, we would obtain less useful spectrum estimates, damaging any further inference. Figure \ref{fig.GaitGroupPSD} displays
an example interval of consecutive PD gait where the gait pattern changed abruptly within the interval. We show that the Welch PSD associated with each of the two gait patterns, varies significantly from the Welch
PSD estimated when grouping both gait patterns together.
An additional problem that arises with estimating Fourier features in free living is that the accurate estimation of the spectrum at the gait frequencies, rests on the assumption of \textit{periodic continuation} \cite{little2019machine}. Because of common non-stationarities in free-living (e.g. mentioned changes within gait episodes, but also the start and end of gait episodes), violations of this assumption are common and can lead to spurious spectral artifacts,
for example caused by Gibbs phenomenon \cite{little2019machine}.
Typically, these issues are ameliorated by using other window
functions than the rectangular window, such as the Hanning window
\cite{allen1977short}. However, while windowing matches samples at window edges (by zeroing), it
also distorts the waveform because it causes \textit{amplitude modulation}. 
In conclusion, the usefulness of spectral estimates largely depends upon accurately locating stationary segments in time, i.e by accurately detecting the start and end of gait episodes, and by detecting (abrupt) changes within gait episodes. At the same time, doing this depends on having access to spectral estimates. Because of this interdependence, we propose a unified framework that addresses both these problems simultaneously.

\begin{figure*}[htbp]  
\center
\includegraphics[width=0.90\columnwidth]{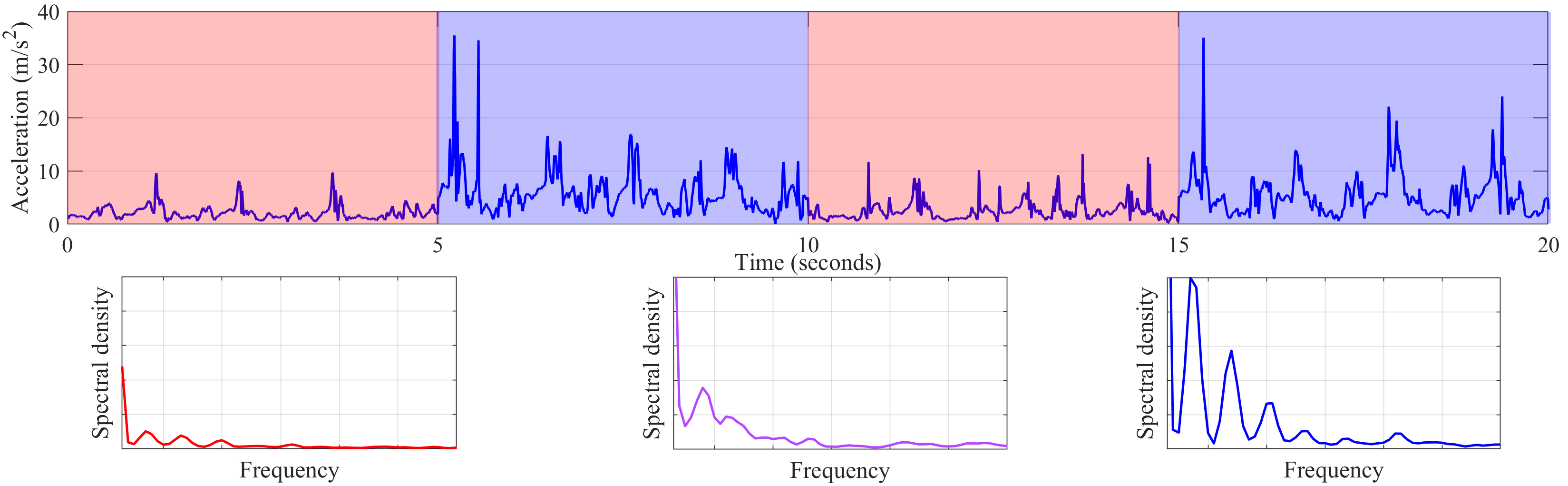}

\caption{Illustrative example of estimating the power spectral density over an unscripted gait segment that contains switches between two different gait patterns (i.e. is approximately piece-wise stationary). The top panel displays the signal magnitude of 20 seconds of pre-processed gait data from a PD patient, obtained from a smartphone worn in the trouser pocket. The red and blue shading indicates different gait states (as identified by the AR-iHMM). The bottom panels display the Welch's PSD estimates: for data from the red gait state (left); for data from the blue gait state (right); for an equal amount of data from both (middle). To allow for same resolution in the 3 bottom plots, we have used 20 seconds of data for each plot.} 
\label{fig.GaitGroupPSD} 
\end{figure*}

\section{Probabilistic modelling of gait \label{sec:Probabilistic-modelling-of gait}}

Most systems focus on segmenting accelerometer data into gait vs.
non-gait classes, by contrast we first segment the data into multiple different groups (more than two) and afterwards assign these groups to
gait or non-gait class. We do this efficiently by
designing a flexible, probabilistic model which is trained directly
on the magnitude of acceleration obtained after removing piecewise
linear device orientation changes (see Section \ref{sec:Challenges of modelling gait}). The proposed model does not rely on any labels, hence it is more objective than a supervised classification approach and reduces the risks of over-fitting.

\paragraph{Autoregressive modelling of gait}

The first assumption we make is that the repetitiveness of the gait
cycle (heel strike, midstance, heel off, midswing, heel strike) is
one of the key properties that characterize gait episodes. The periodic
nature of the accelerometer data during gait \cite{moe2004estimation}
makes it efficient to detect and model gait based on the spectrum,
for example using the Fourier transform. However, Fourier spectral
analysis inherently assumes periodic continuation (see Section \ref{sec:Challenges of modelling gait}). We address this problem by
simultaneously estimating the spectrum and the start and end points
of the stationary gait episodes. To achieve this, we first model the
spectrum of the gait in the time domain, using an \textit{autoregressive
(AR) processes} \cite{little2019machine}. An order $r$ AR model is a random process which describes a sequence $\left\{ x_{t}\right\} _{t=1}^{T}$ as a linear combination of previous values in the sequence and a stochastic term:
\begin{equation}
x_{t}=\sum_{j=1}^{r}A_{j}x_{t-j}+e_{t}\qquad e_{t}\sim\mathcal{N}\left(0,\sigma^{2}\right)\label{eq:Autoregressive process-1}
\end{equation}
where $A_{1},\dots,A_{r}$ are the AR coefficients, $T$ denotes the
length of the sequence $e_{t}$ is a zero mean random variable, assumed to be an i.i.d. Gaussian sequence (we can trivially extend the model such that $e_{t}\sim\mathcal{N}\left(\mu,\sigma^{2}\right)$
for any real-valued $\mu$). We assume that the AR noise variance
$\sigma^{2}$ is unknown and place a conjugate \textit{inverse-Wishart} prior over it. This essentially means that in addition to modelling
the periodicity of the input signals, we also account for changes in the 
non-periodic components of the signals. We saw in Section \ref{sec:Challenges of modelling gait} that the window variance of the acceleration can be a useful discriminator of gait versus non-gait on its own in certain scenarios. If we assume an AR model of order $r=0$, the variance of $e_{t}$ is the variance of the window.

AR processes are commonly used as parametric models of the PSD since
the power spectrum is determined by the AR parameters \cite{little2019machine}: 
\begin{equation}
S\left(f\right)=\frac{\sigma^{2}}{\left|1-\sum_{j=1}^{r}A_{j}\exp\left(-i2\pi fj\right)\right|^{2}}
\end{equation}
where $f$ is the frequency variable and
$i$ denotes the imaginary unit. This means that the number
of non-zero AR coefficients determines the complexity of the PSD which the  model can represent: there is a peak in the PSD for each complex-conjugate pair of roots of the coefficient polynomial. Parametric
spectral estimation is often more stable than non-parametric PSD methods,
and can be of high quality using fairly little data, assuming the model is correct. By contrast, non-parametric
PSD estimation methods require more data to produce stable estimates, unless we trade some of the frequency resolution via averaging,
as in Welch's PSD \cite{little2019machine}.
The parametric model of the spectrum will allow us to construct a
flexible, non-parametric model of the switching dynamics of different
gait and non-gait activities in free living. More detailed discussion
on the relative merits of different spectral estimation methods combined
with machine learning, can be found in Little \cite{little2019machine}.

\paragraph{High order adaptive autoregressive processes}

As mentioned above, the AR order $r$ we use will
determine the complexity of this parametric model of the spectrum.
The optimal AR model $r$ is likely to vary across different stationary
segments of sensor data and choosing fixed $r$ which is too large
will lead to problems with parameter estimation (fitting the AR coefficients). At the same time, gait is typically characterized by a low fundamental frequency, with bandwidth of up to 10-15Hz (see Section \ref{sec:Related-work}).
This implies the need for fairly high order $r$ AR processes (together
with sufficiently high sample rate) in order to accurately capture
the typical range of gait frequencies. To address this conflict, we use a non-conjugate Bayesian prior on the AR coefficients $A_{1},\dots,A_{r}$
which induces \emph{sparsity} of the coefficients (only a few are
non-zero at any one time) and allows us to draw conclusions about the
autoregressive model coefficients that do not contribute to the underlying
dynamics of the gait. In effect, this means that we attempt to learn
fewer than $r$ AR coefficients supported by the signal
but potentially associated with larger AR time delays. This is done
by assuming independent, zero-mean Gaussian priors on the coefficients
$A_{1},\dots,A_{r}$ with unknown precisions, which acts as an \textit{automatic
relevance determination prior }(ARD) \cite{Mackay1995}. The ARD
prior was first proposed in the context of neural network models in
Mackay \cite{Mackay1995} and then later adopted for discrete latent variable
models in Beal \cite{beal2003variational} and for switching AR processes
in Fox et al. \cite{fox2009bayesian}.

\begin{table*}
\caption{Gait detection performance of the proposed AR-iHMM and of common gait detection algorithms (using the thresholds reported in the literature, and after pre-processing and optimizing thresholds). We have computed the average performance and standard deviation using leave-one-subject-out cross-validation. For PD patients, we show the performance of the complete free-living assessments, and the difference in balanced accuracy (average of sensitivity and specificity) between the parts before and after medication intake. \label{tab:Gait-detection-performance-thresholds}}

\center
\begin{tabular}{*6c}
\Xhline{2\arrayrulewidth}
Name & \multicolumn{3}{c}{PD} & \multicolumn{2}{c}{Control}  \tabularnewline
\Xhline{2\arrayrulewidth}
 & Sensitivity & Specificity & \begin{tabular}{@{}c@{}}Accuracy difference \\ before/after medication\end{tabular} & Sensitivity & Specificity \tabularnewline
\hline
\hline
\multicolumn{6}{c}{ \textit{With fixed thresholds}\qquad \qquad \qquad \qquad \qquad \qquad \qquad \qquad \qquad \qquad \qquad \qquad \qquad \qquad \qquad \qquad \qquad \qquad } \tabularnewline
\hline
\hline
STD-thresholding & 77\% (15\%) & 93\% (4\%) & 5\% & 90\% (6\%) & 96\% (4\%) \tabularnewline
\hline 
STFT-thresholding & 85\% (12\%) & 94\% (4\%) & 3\% & 86\% (6\%) & 85\% (6\%)   \tabularnewline
\hline 
NASC  & 87\% (9\%) & 96\% (4\%) & 3\% & 88\% (6\%) & 97\% (4\%)  \tabularnewline
\hline 
CWT-thresholding  & 66\% (10\%) & 97\% (5\%) & 5\% & 70\% (4\%) & 97\% (4\%)  \tabularnewline
\hline 
\hline
\multicolumn{6}{c}{ \textit{With optimized thresholds and pre-processing}\qquad \qquad \qquad \qquad \qquad \qquad \qquad \qquad \qquad \qquad \qquad \qquad \qquad} \tabularnewline
\hline
\hline
STD-thresholding & 90\% (14\%) & 91\% (4\%) & 5\% & 93\% (5\%) & 92\% (3\%)  \tabularnewline
\hline 
STFT-thresholding & 91\% (11\%) & 89\% (4\%) & 3\% & 92\% (5\%) & 91\% (4\%)   \tabularnewline
\hline 
NASC  & 93\% (10\%) & 87\% (5\%) & 3\% & 94\% (5\%) & 91\% (3\%)  \tabularnewline
\hline 
CWT-thresholding  & 92\% (8\%) & 85\% (5\%) & 5\% & 94\% (3\%) & 85\% (6\%)  \tabularnewline
\Xhline{2\arrayrulewidth}
\begin{tabular}{@{}c@{}}\textbf{Probabilistic modelling} \\ \textbf{(based on our AR-iHMM)}\end{tabular} & \textbf{91\% (8\%)} & \textbf{91\% (3\%)} & \textbf{1\%} & \textbf{94\% (5\%)} & \textbf{93\% (3\%)}  \\
\Xhline{2\arrayrulewidth}
\end{tabular}
\end{table*}
\paragraph{Latent switching behavior dynamics}

To analyze free living data, it is insufficient to define a
parametric spectral model for the patients' gait, because
participants regularly switch between different gait and non gait episodes,
which results in highly non-stationary time series (see Figure \ref{fig.AccelerometerData}
and Figure \ref{fig.GaitGroupPSD}).

Even within gait episodes, the optimal AR parameters to model the
gait might change depending on the speed, amplitude and other characteristics
of the walking pattern. In order to group similar gait signals, but
also separate gait from non-gait data, we use a \textit{switching AR process} model \cite{Kim1994} (AR-HMM). However, one drawback of conventional switching AR processes,
is that it requires a fixed number of hidden states and AR order. Since the heterogeneity in both
gait and non-gait episodes will increase as more free living data
becomes available, we adapt the more flexible \textit{non-parametric
switching AR process} first proposed in Fox et al. \cite{fox2009nonparametric}.
The model can be thought of as an infinite-state extension of the
model of Kim \cite{Kim1994} (AR-iHMM).
Viewing the switching AR model as \textit{\emph{a hidden Markov model}}
(HMM) with AR processes used to model the HMM emissions, then in the
non-parametric switching AR model the parametric HMM is effectively
replaced with an \emph{infinite} HMM \cite{Beal2002}.

In the AR-iHMM model, we assume that the data is an inhomogeneous
stochastic process and that multiple AR models are required to represent
the dynamic structure of the signal, i.e.:
\begin{equation}
x_{t}=\sum_{j=1}^{r}A_{j}^{z_{t}}x_{t-j}+e_{t}^{z_{t}}\qquad e_{t}^{z_{t}}\sim\mathcal{N}\left(0,\sigma_{z_{t}}^{2}\right)
\end{equation}
where $z_{t}\in\left\{ 1,\dots,K^{+}\right\} $ indicates the AR model
associated with time index $t$. The latent variables $z_{1},\dots,z_{T}$
describing the switching process are modelled with a Markov chain.
A transition matrix $\pi$ is estimated with $K^{+}$ rows and $K^{+}+1$
columns indicating the probability of specific transitions from existing
state $i$ to existing state $j$, $\pi_{ij}$, or from existing state
$i$ to a new state $K^{+}+1$, $\pi_{iK^{+}+1}$. Transitions that
are observed more often during the training of the model will have
higher probability, represented in the transition term $\pi_{ij}$.

When $K^{+}\ll T$, this model clusters together parts of the signal
into an, a priori, unknown number $K^{+}$ of time segments which
are best represented with the same AR coefficients. In AR-iHMM, $K^{+}$
is unknown: instead of being fixed it is inferred from the data and
can adapt to new, unseen structure in the data. The AR-iHMM
is obtained by augmenting the transition matrix of the Markov process
$\pi$ underlying the latent variables $z_{1},\dots,z_{T}$ with a
\emph{hierarchical Dirichlet process} (HDP) \cite{Teh2004} prior.

\section{Empirical comparison of gait detection algorithms\label{sec:Comparison-with-gait}}

In order to make inferences about the gait pattern, we first need
to verify that our proposed framework is able to identify gait segments.
In this section, we evaluate our ability to detect gait as annotated
in the video recordings of the Parkinson@Home validation study. To
establish whether our approach achieves reasonable results, we include
a comparison with widely used gait detection algorithms. It is important
to note that our goal was not to maximize gait detection accuracy (when compared with human annotation),
but to locate gait segments in time, that are useful to study the effect of PD on the gait pattern.

In this section, we use the pre-processed accelerometer data (see Section \ref{sec:Challenges of modelling gait}) from the smartphones and Physilog 4 devices placed on various body locations (see Section \ref{sec:Data description}). We infer the AR-iHMM described in Section
\ref{sec:Probabilistic-modelling-of gait} using scalable iterative MAP inference proposed in Raykov et al. \cite{Raykov2016}. Any hyperparameters associated with the AR state priors or the HDP prior (see Section \ref{sec:Probabilistic-modelling-of gait}) are fixed across patients and are selected using standard Bayesian model selection. For each point $x_t$, we consider it is associated with its most likely state $z_t = k*$ to enable direct comparison, i.e. we ignore the estimated uncertainty associated with the segmentation indicators. 

To determine if the identified hidden Markov states should be classified as gait or non-gait,
we consider the AR-based PSD estimates associated with each state. Specifically, we compute the total energy at frequencies in the range {[}0.5 - 10Hz{]}, and select a threshold of minimal spectral energy that maximizes the balanced accuracy (average of sensitivity and specificity) averaged across participants (measured against the manual video annotations for the presence of gait). We evaluate the performance of selecting the threshold using leave-one-subject-out cross-validation. Thresholding using a shared PSD range across participants is done only to enable a fair and intuitive comparison with the other commonly used techniques for detection of gait in smartphones and wearables; in principle, once the AR-iHMM model is trained we can derive multiple features related to the distribution of the sensor data and train a supervised classifier on these features.  

\begin{figure}[htbp]  
\centering
\includegraphics[width=0.45\columnwidth]{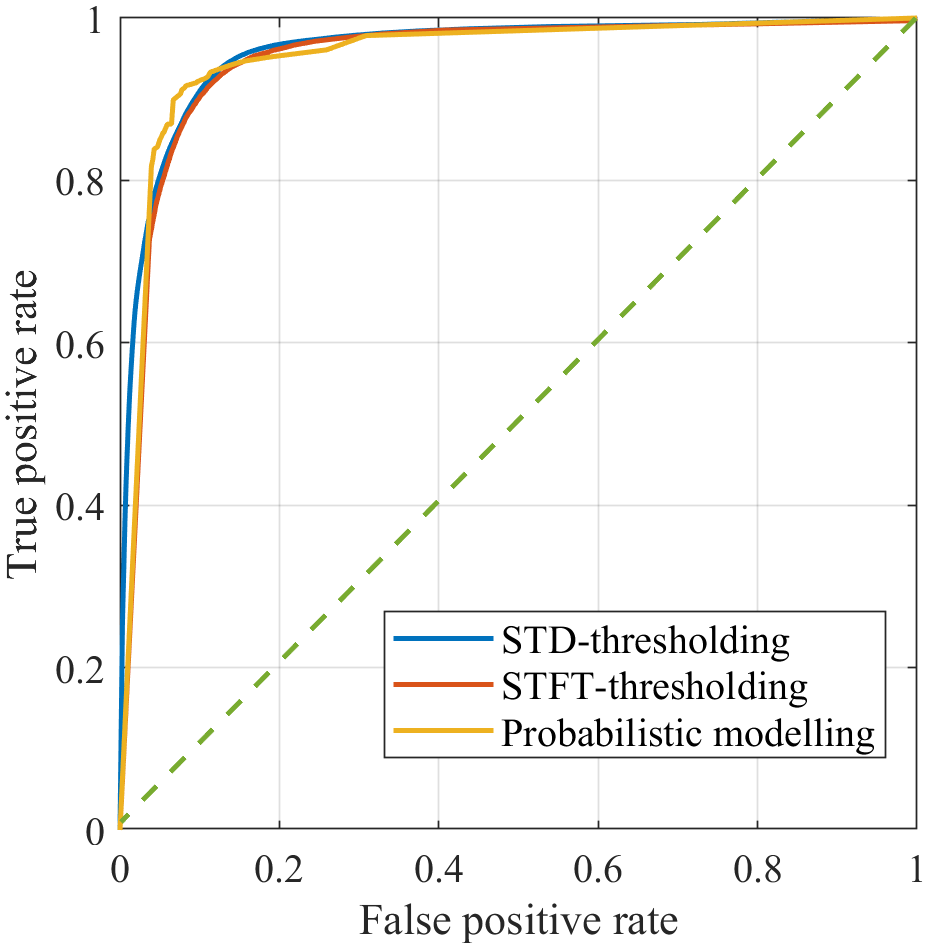}
\caption{ROC curve of different methods used for gait detection. To obtain the ROC of the probabilistic modelling (AR-iHMM), we have varied the threshold of the spectral energy in range [0.5-10Hz], obtained after estimating the AR-iHMM; varying different model parameters can lead to different curves. For the remaining thresholding methods, we have varied the single threshold. In green is the performance of a random classifier.} 
\label{fig.ROC} 
\end{figure}

For the comparison with existing algorithms, we implemented the following widely used generic gait detection algorithms: STD-thresholding \cite{Randell2003,brajdic2013};
STFT-thresholding \cite{Allen1977}; normalized autocorrelation step
detection and counting (NASC) \cite{Rai2012} and continuous wavelet
transform (CWT) thresholding \cite{barralon2006walk}.We evaluate the performance of the original formulations of the algorithms, and the performance after applying our pre-processing pipeline and adjusting corresponding thresholds to maximize the balanced accuracy across participants using leave-one-subject-out cross-validation\footnote{Different thresholds are used for PD and controls cohorts.}:
\begin{itemize}
\item STD-thresholding: we set a threshold based on the 1 second window
standard deviation which optimally discriminates gait from non-gait classes;
\item STFT-thresholding: we set a threshold based on the 1 second window
total energy at frequencies in the range {[}0.5 - 10Hz{]} to maximize the balanced accuracy averaged across
participants;
\item NASC algorithm: the NASC involves first applying STD-thresholding
and then evaluating the auto-correlation of the remaining data over
2 second windows, specifically looking at the time delays representative
of gait. We set a modified STD threshold, a range of delays, and an
auto-correlation threshold to maximize the balanced accuracy averaged
across participants (iteratively, one at a time);
\item CWT-thresholding: we compute the ratio between the energy in the band
of walking frequencies and the total energy across all frequencies,
and set a threshold to maximize the balanced accuracy averaged across
participants.
\end{itemize}
In Table \ref{tab:Gait-detection-performance-thresholds} we report the performance of the different methods when applied to the smartphone data. In Figure \ref{fig.ROC}, we plot the trade-off between sensitivity and specificity as we vary the threshold in the receiver operating characteristic (ROC) curve for the different methods. Finally, in Table \ref{tab:Gait-detection-accuracy locations} we evaluate all algorithms on data from different accelerometry devices, placed at different body locations (see Section \ref{sec:Data description}).

\subsection{Inclusion criteria\label{sec:Inclusion criteria}}
It should be noted that we do not provide a comprehensive comparison
of all available gait detection algorithms: we have omitted some because (1) they were largely based on
heuristics which could not be trivially adapted for detection of pathological
gait; (2) they demonstrated very poor performance on our dataset;
or (3) they had strong conceptual overlap with the techniques included
in the comparison. In addition, we have excluded activity recognition
pipelines that rely on a large number of features to detect gait.
The main reason for this is the inherent curse of dimensionality which
we want to account for in most health monitoring applications. The
huge variation in unconstrained free living data, coupled with the
fairly limited amount of labeled data (in the Parkinson@home validation
study a few hours from 50 participants) suggest that there is high
risk of overfitting our training data and collecting unreliable gait
measurements out-of-sample in the unsupervised setup. Because of this,
we propose that one should aim for principled gait detection criteria,
rather than simply maximizing the accuracy on training datasets.

\begin{table*}
\caption{Performance of different methods for gait detection across different sensor locations (after pre-processing and optimizing thresholds). The performance is expressed as balanced accuracy (average of sensitivity and specificity), evaluated on the complete free living assessments of PD patients against video annotations. \label{tab:Gait-detection-accuracy locations}}

\centering{}%
\begin{tabular}{*7c}
\Xhline{2\arrayrulewidth}
Name & Left Shin & Right Shin & Left Wrist & Right Wrist & Lower Back & Smartphone\tabularnewline
\Xhline{2\arrayrulewidth}

STD-thresholding & 94\% & 94\% & 77\% & 76\% & 91\% & 90\%\tabularnewline
\hline 
STFT-thresholding & 93\% & 92\% & 83\% & 81\% & 90\% & 90\%\tabularnewline
\hline 
NASC & 91\% & 91\% & 78\% & 77\% & 92\% & 90\%\tabularnewline
\hline 
CWT-thresholding & 92\% & 90\% & 75\% & 72\% & 90\% & 89\%\tabularnewline
\Xhline{2\arrayrulewidth}
\textbf{Probabilistic modelling} & \textbf{94\%} & \textbf{94\%} & \textbf{83\%} & \textbf{83\%} & \textbf{93\%} & \textbf{91\%}\tabularnewline
\Xhline{2\arrayrulewidth}
\end{tabular}
\end{table*}

\subsection{Discussion}
First of all, the results in Table \ref{tab:Gait-detection-accuracy locations} show that it is feasible to identify gait using our modelling approach, with at least as good average performance compared to existing algorithms. In addition, the results  underline the importance of appropriate pre-processing and threshold adjustment when applying algorithms to patients with PD. 

In most methods, we observe a difference in accuracy between PD patients and controls, and between before and after medication intake for PD patients.  The latter is most notable in patients with a strong response to medication. This difference in accuracy between before and after medication intake was less prominent for the AR-iHMM, which also demonstrated less variability in the performance across PD patients. Moreover, the performance of the AR-iHMM was relatively robust to different body locations of the sensor in comparison to STD-thresholding, NASC, and CWT-thresholding (Table \ref{tab:Gait-detection-accuracy locations}). While balanced accuracy is fairly similar, the AR-iHMM consistently captures longer gait segments, but misses on shorter duration walks, whereas the remaining techniques are more accurate on shorter periods, but interrupt long consistent walking intervals.

It is worth noting that the occurrences (prevalence) of the gait/non-gait classes are not balanced in the free living assessments from the Parkinson@Home validation study. Across PD patients, the mean walking time is 16\% with a standard deviation of 6\%. The mean walking time is slightly higher for non-PD controls at 21\% with the same standard deviation. Because we cannot assume that this prevalence is representative of truly free living situations, we choose to evaluate the methods with measures that are independent of the prevalence of gait (i.e. sensitivity and specificity). However, the thresholds were set to optimize the balanced accuracy (mean of sensitivity and specificity), which implicitly optimizes for the situation where class prevalences are equal, and misclassification costs are equal as well. Different applications may require different settings of the thresholds.

In contrast with the other methods evaluated, the trained AR-iHMM model offers a simplified representation of the data, which can be used
to distinguish between multiple gait and non-gait patterns. We discuss the use of the different gait states in the next Section \ref{sec:Modelling-gait-pattern}.

\section{Modelling gait pattern changes \label{sec:Modelling-gait-pattern}}

The non-parametric nature of our approach allows us to rigorously identify 
different segments in a data driven manner. In addition to gait detection,
this segmentation can also be used to identify clinical changes in
the gait pattern. We demonstrate this in PD patients by showing that
the gait segments identified by our algorithm, have a different gait pattern before
and after intake of symptomatic medication.

\paragraph{Discrimination of gait before/after medication intake}

An important potential application of free living gait analysis in
PD patients is monitoring real-life variations in the response to
medication. It is well known that dopaminergic medication (e.g. levodopa)
can have a visible effect on PD patients' ability to walk \cite{smulders2016pharmacological}.
Here we investigate the effectiveness of our probabilistic modelling
approach for the problem of classifying gait episodes into ``before
medication'' and ``after medication'' classes. The comparison
is done using smartphone accelerometer data from the home visits of
the Parkinson@Home validation study (described in Section \ref{sec:Data description}).

For this classification problem, we consider all segments that were identified
as gait by our model (see Section Empirical comparison of gait detection
algorithms'). The AR-iHMM segments the data into intervals with the
state variables $z_{t}$ denoting the AR state representing the
signal at time indexed $t$. If we then assume $K^{+}$ unique values
for $z_{t}$ as $t=1,\dots,T$, we will estimate $K^{+}$ sets of
AR coefficients: $\left\{ A_{1}^{k},\dots,A_{r}^{k}\right\} _{k=1}^{K^{+}}$.
For each state $k$ we estimate the spectrum based on AR coefficients
$A_{1}^{k},\dots,A_{r}^{k}$.

\begin{figure}[t]  
\centering
\includegraphics[width=0.495\columnwidth]{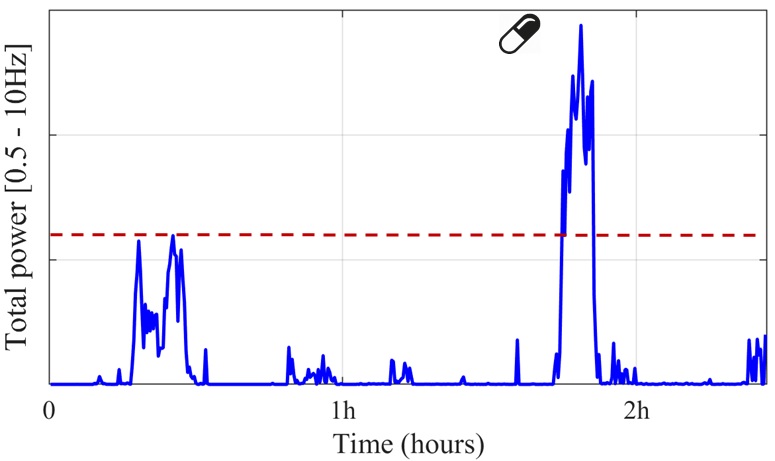}
\includegraphics[width=0.495\columnwidth]{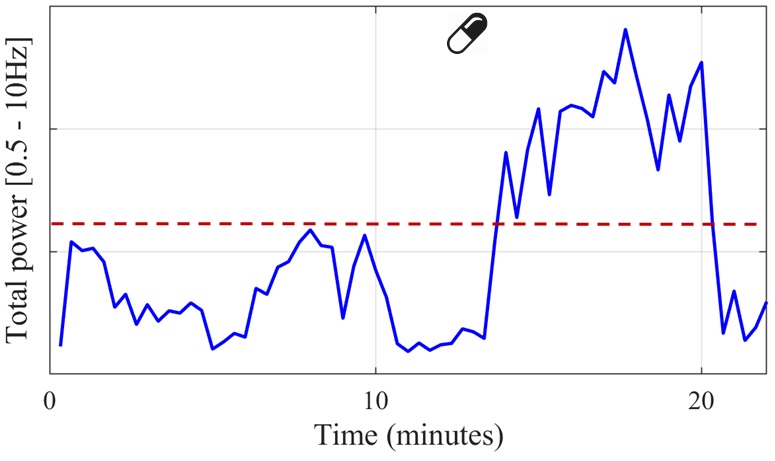}
\caption{Left: the total power in the 0.5-10Hz range, obtained from the smartphone, during the complete free living assessment of one PD patient that shows a clear change after medication intake. Right: the total power in the 0.5-10Hz during gait as identified by the AR-iHMM. The pill icon indicates the time of medication intake (~40 minutes of data directly after medication intake have been cropped, because this contained the standardized clinical assessment) The horizontal red line indicates a viable decision boundary between gait occurring before and after medication intake in this patient.} 
\label{fig.TotalEnergy0.5_10Hz} 
\end{figure}

There are multiple PSD features that in principle can be used to monitor
PD related changes in the gait pattern: position of the dominant peak
(often the fundamental frequency); height of the dominant peak; width
of the dominant peak; ratio of the first and second peak; energy
in a specific frequency range, and others \cite{del2016}. In our
evaluation, we consider the height and position of the dominant peak,
and the total energy in the range 0.5-10Hz (gait related information
is expected in this frequency range). In Figure \ref{fig.TotalEnergy0.5_10Hz}
we illustrate how the total energy estimated from the PSD varies throughout
a single visit and how it varies during the predicted gait periods
of that visit. Because we expect that the relative rather than the
absolute within-person changes are relevant to distinguish between before and after medication intake, we normalize
all features per patient using z-scores. 

Of the 25 PD patients taking part in the
study, 18 were considered to have sufficient walking periods both before and
after medication. In these patients, we train a logistic classifier using each of the features
described above, to predict whether a gait segment occurred before or after medication intake. For each patient, we compute the out-of-sample
accuracy based on leave-one-subject-out cross validation\footnote{The accuracy of the leave-one-subject-out cross validation is affected
both by the flexibility of the trained classifier, but also the intrinsic
variation across features from different subjects.}. As displayed in Table \ref{tab:Average-accuracy}, we can predict with reasonable accuracy whether a gait segment occurred before or after medication intake; not all patients have visible motor response after medication
intake, so we expect that achieving perfect prediction accuracy will
not be possible. Note that combining multiple gait features may slightly increase accuracy, but the goal here is to obtain an interpretable model.

\begin{table}
\caption{Balanced accuracy (average of sensitivity and specificity) to predict whether gait segments occurred before or after medication intake, using a logistic classifier based on PSD features obtained from the AR-iHMM, normalized per subject. We use leave-one-subject-out cross validation, and present the mean and standard error across subjects. Results are compared between using gait segments as annotated on the video recordings (''Annotated gait''), and gait segments identified by our model (''Predicted gait''). \label{tab:Average-accuracy}}

\centering{}%
\begin{tabular}{*3c}
\Xhline{2\arrayrulewidth}
Feature & Annotated gait & Predicted gait\tabularnewline
\Xhline{2\arrayrulewidth}
Peak hight & 72\% (9\%) & 70\% (10\%)\tabularnewline
\hline 
Peak position & 51\% (14\%) & 50\% (18\%)\tabularnewline
\hline 
Total energy & 74\%(8\%) & 78\% (8\%)\tabularnewline
\Xhline{2\arrayrulewidth}
\end{tabular}
\end{table}

To examine how our approach for gait detection affects the ability to discriminate between before and after medication intake, we compare results with using the gait segments as annotated on the video recordings. Here, we learn the AR-iHMM on all annotated gait data, and use all the identified states to obtain the AR-based PSD. We see that the accuracy when using the gait segments identified by our model is at least as high as using the annotated gait segments. When using the total energy, the model-based gait segments appear to be even more informative. A possible explanation for this is that not all gait segment are equally informative, and that the model selects longer, more periodic, ''steady-state'' gait segments, which are less sensitive to environmental or behavioral factors that are irrelevant for the classification problem.

\paragraph{Exploratory gait analysis}

\begin{figure*}[htbp]  
\centering
\includegraphics[width=0.95\columnwidth]{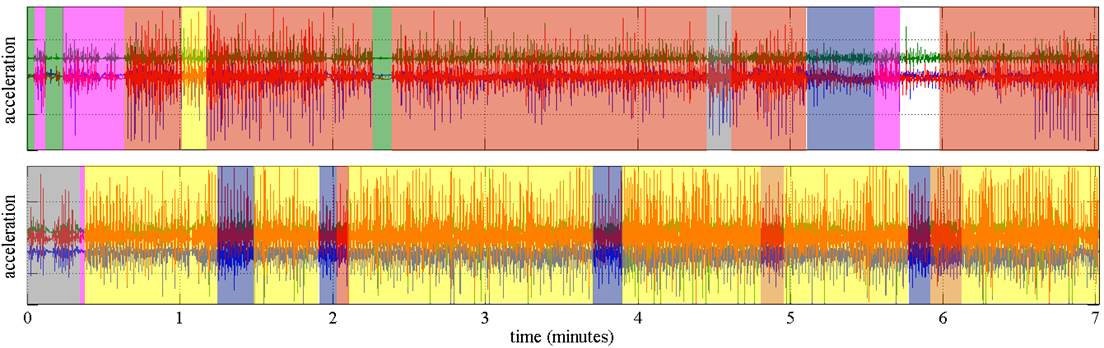}
\caption{Segmentation of smartphone data during the free living assessment obtained from the AR-iHMM, from one PD patient with clinically observable changes in the gait pattern after medication intake (based on the video recordings). Top: before medication intake. Bottom: after medication intake. The red, yellow, blue and grey segments are all associated with gait data (according to both the video annotations and the classification based on the AR-iHMM-based PSD features); the remaining segments indicate different non-gait patterns.} 
\label{fig.ClusteringGait} 
\end{figure*}

\begin{figure*}[htbp]  
\centering
\includegraphics[width=0.95\columnwidth]{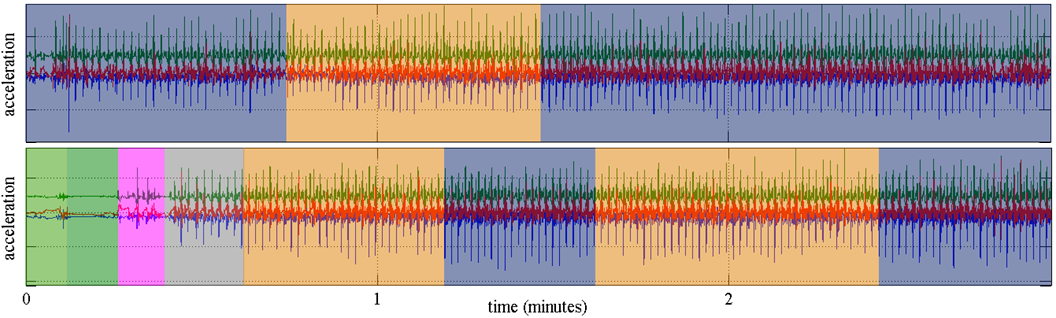}
\caption{Segmentation of smartphone data during the free living assessment obtained from the AR-iHMM, from one PD patient with no clinically observable changes in the gait pattern after medication intake (based on the video recordings). Top: before medication intake. Bottom: after medication intake. The yellow, blue and grey segments are all associated with gait data, with similar occurrence before and after medication intake; the remaining segments are not associated with gait.} 
\label{fig.ClusteringGaitNonResponsive} 
\end{figure*}

Because our probabilistic model is unsupervised, we can use the model not only
as a tool to make predictions, but also as an exploratory tool to study
the gait data. For example, in Figure \ref{fig.ClusteringGait} we
show the gait segmentation of a PD patient with a notable clinical
improvement in gait pattern after medication intake (based on
the video annotations), where different colors indicate different hidden states
$z_{t}$. What we observe is that the probabilistic model not only
allows us to identify non-gait segments (pink and green), but
also discriminates between different variations in gait quality. In this
patient, the model separates before medication gait (red) or after medication gait (yellow). By contrast, Figure \ref{fig.ClusteringGaitNonResponsive} shows segmented gait
of a PD patient whose gait does not notably improve after medication
intake (based on the video recordings). Interestingly, we can still
identify different gait segments both before and after medication
intake, but their pattern of occurrence is similar in both conditions.
Furthermore, inspection of the AR-based PSD estimates associated with
the states in both figures indicates that the gait states in Figure
\ref{fig.ClusteringGaitNonResponsive} are more similar to each other
than the gait states associated with before and after medication periods
in Figure \ref{fig.ClusteringGait}. It should be noted that this
contrast was not present in all patients, and we show two illustrative
cases here. Further research is needed to identify reasons why some
cases do not show the expected change.

\section{Discussion\label{sec:Discussion and conclusion}}

In this report we study the problem of passively monitoring movement disorders such
as Parkinson's disease (PD) in daily living using wearable sensors.
This is a challenging problem for two main reasons. First, many factors
other than the changes to the medical condition, 
contribute to the enormous variation seen in daily living sensor signals,
such as voluntary behaviour and device orientation. Second, it is
costly and logistically difficult to collect representative data in
daily living with reliable labels. This may explain why highly flexible
methods such as deep learning have not been successful in the context
of monitoring symptom fluctuations in PD \cite{hammerla2015pd}.
This has stimulated the search for signal models that are based on
principled assumptions which reduce the model's flexibility while
still allowing it to capture subtle disease-related changes. In this work we
propose a simple, structured probabilistic modelling approach specifically
for the analysis of free living gait. Gait is a promising target for
passive monitoring because it is a common behaviour and responds to
symptomatic medication in PD patients. Our approach is designed to
simultaneously locate stationary gait segments and characterize the
gait pattern based on pre-processed accelerometer data. We achieve
this by adopting a non-parametric switching autoregressive model,
circumventing the need to use window-based analysis and the need to pre-define
the number of gait and non-gait classes that can be observed in daily
life.

We demonstrate our approach on a new reference dataset including 25 PD
patients and 25 controls. The dataset is unique because it combines unscripted daily living activities in and around the house with detailed video annotations, which allows us to test our model on a much more realistic setting. First, we show that the identified classes
can be used to accurately detect gait.
Second, we show that states that represent gait can be used to predict
medication induced fluctuations in PD patients.

There are other potential advantages of the proposed model which
are not reflected in empirical classification accuracies. Our approach has two key advantages when it comes to estimating
the spectrum of the free living accelerometer data (or similar sensors): (1) the time boundaries of each segment of stationary
data over which we compute the spectrum are adaptively selected by
the model, which avoids the need for window-based analysis and problems associated with this; (2) in a fully probabilistic fashion, we can leverage multiple
repeating patterns to get a more robust estimate of the spectrum.

Additionally, because
our algorithm does not treat the problem as binary classification (gait/no
gait), but is designed instead to learn multiple gait and non-gait states,
it can deal with changes in gait pattern during gait episodes.
This avoids grouping different gait patterns together, which can introduce
problems in further gait pattern analyses. Moreover, because it is
unsupervised, the model can be used as an exploratory tool
to locate gait or non-gait segments that share the same (spectral)
characteristics. The user can explicitly control the prior parameters
of the model to determine the temporal granularity of the segmentation
and focus in on more or less detailed changes in the gait signals.
This extra control allows us to focus on sufficiently stationary (``steady
state'') gait segments of good quality gait data representative of
patient's PD symptoms. Lastly, using this fully Bayesian model to describe
the acceleration signals allows us to estimate the uncertainty involved
with both the segmentation and the estimation of the spectrum.

\subsection{Limitations and future directions}

In order to develop an intuitive, robust and easy to interpret probabilistic signal model for gait data in free living, we have made some restrictive assumptions about the distribution and occurrence of such data in daily life. Despite the flexibility of inferring an unknown number of different spectral AR representations, we have focused on the states that have sufficient spectral power in the gait range to monitor changes after medication intake. We believe this approach is appropriate for monitoring the highly prevalent continuous impairments in PD patients (\textit{bradykinetic gait}). However, by focusing on short-term interruptions of the gait, the model could potentially also be very suitable to monitor the more rare episodic hesitations (\textit{freezing of gait}). Because of the limited number of patients that presented with this symptom during the free-living assessments of the Parkinson@Home validation study, this remains to be evaluated using other data sets. In addition, the proposed framework does not use the axis meanings in the sensor outputs. This was done because in smartphones, the default orientation of the device can be different depending on the user. Our approach can also be applied to the three-dimensional dynamic component of the acceleration vector, or to any specific axis.
Before the framework can used in medical contexts, further validation is necessary. For example, in the current study protocol the gait before medication intake was measured after overnight withdrawal of dopaminergic medication. While this allowed for a detailed assessment of the changes after medication intake, in some patients the effects in daily life might be more subtle. Future work will aim to evaluate how well response fluctuations can be captured for naturally occurring shorter withdrawal periods in truly unsupervised conditions. 
Finally, we emphasize that the developed framework aims to merely segment varying gait patterns in a principled and largely unsupervised manner. In order to assign meaning to detected changes in the gait pattern observed in real-life, e.g. estimate causal effects of medication, further analysis using a carefully designed causal map is required. For example, real-life factors such as environment (e.g. crowded city versus park) might also influence the observed gait pattern. Future work will include adjusting for real-life confounding, using contextual sensors such as GPS.

\bibliographystyle{IEEEtran}
\bibliography{references4.bib}

\begin{thebibliography}{10}
\providecommand{\url}[1]{#1}
\csname url@samestyle\endcsname
\providecommand{\newblock}{\relax}
\providecommand{\bibinfo}[2]{#2}
\providecommand{\BIBentrySTDinterwordspacing}{\spaceskip=0pt\relax}
\providecommand{\BIBentryALTinterwordstretchfactor}{4}
\providecommand{\BIBentryALTinterwordspacing}{\spaceskip=\fontdimen2\font plus
\BIBentryALTinterwordstretchfactor\fontdimen3\font minus
  \fontdimen4\font\relax}
\providecommand{\BIBforeignlanguage}[2]{{%
\expandafter\ifx\csname l@#1\endcsname\relax
\typeout{** WARNING: IEEEtran.bst: No hyphenation pattern has been}%
\typeout{** loaded for the language `#1'. Using the pattern for}%
\typeout{** the default language instead.}%
\else
\language=\csname l@#1\endcsname
\fi
#2}}
\providecommand{\BIBdecl}{\relax}
\BIBdecl

\bibitem{Lipsmeier2017}
F.~Lipsmeier, I.~Fernandez~Garcia, D.~Wolf, T.~Kilchenmann, A.~Scotland,
  J.~Schjodt-Eriksen, W.-Y. Cheng, J.~Siebourg-Polster, L.~Jin, J.~Soto,
  L.~Verselis, M.~Martin~Facklam, F.~Boess, M.~Koller, M.~Grundman, M.~A.
  Little, A.~Monsch, R.~Postuma, A.~Gosh, T.~Kremer, K.~Taylor, C.~Czech,
  C.~Gossens, and M.~Lindemann, ``Successful passive monitoring of early-stage
  {P}arkinson's disease patient mobility in {P}hase {I RG7935/PRX002} clinical
  trial with smartphone sensors,'' \emph{Movement Disorders}, vol.~32, no.~2,
  2017.

\bibitem{sha2008spa}
K.~Sha, G.~Zhan, W.~Shi, M.~Lumley, C.~Wiholm, and B.~Arnetz, ``{SPA}: a smart
  phone assisted chronic illness self-management system with participatory
  sensing,'' in \emph{Proceedings of the 2nd International Workshop on Systems
  and Networking Support for Health Care and Assisted Living Environments},
  2008, pp. 5:1--5:3.

\bibitem{horak2013objective}
F.~B. Horak and M.~Mancini, ``Objective biomarkers of balance and gait for
  parkinson's disease using body-worn sensors,'' \emph{Movement Disorders},
  vol.~28, no.~11, pp. 1544--1551, 2013.

\bibitem{warmerdam2020long}
E.~Warmerdam, J.~M. Hausdorff, A.~Atrsaei, Y.~Zhou, A.~Mirelman, K.~Aminian,
  A.~J. Espay, C.~Hansen, L.~J. Evers, A.~Keller \emph{et~al.}, ``Long-term
  unsupervised mobility assessment in movement disorders,'' \emph{The Lancet
  Neurology}, 2020.

\bibitem{De2016}
A.~L.~S. de~Lima, T.~Hahn, N.~M. de~Vries, E.~Cohen, L.~Bataille, M.~A. Little,
  H.~Baldus, B.~R. Bloem, and M.~J. Faber, ``Large-scale wearable sensor
  deployment in parkinson’s patients: the parkinson@ home study protocol,''
  \emph{JMIR research protocols}, vol.~5, no.~3, 2016.

\bibitem{arora2014high}
S.~Arora, V.~Venkataraman, S.~Donohue, K.~M. Biglan, E.~R. Dorsey, and M.~A.
  Little, ``High accuracy discrimination of {P}arkinson's disease participants
  from healthy controls using smartphones,'' in \emph{2014 IEEE International
  Conference on Acoustics, Speech and Signal Processing (ICASSP)}, 2014, pp.
  3641--3644.

\bibitem{bot2016mpower}
B.~M. Bot, C.~Suver, E.~C. Neto, M.~Kellen, A.~Klein, C.~Bare, M.~Doerr,
  A.~Pratap, J.~Wilbanks, E.~R. Dorsey \emph{et~al.}, ``The mpower study,
  parkinson disease mobile data collected using researchkit,'' \emph{Scientific
  data}, vol.~3, p. 160011, 2016.

\bibitem{del2019gait}
S.~Del~Din, M.~Elshehabi, B.~Galna, M.~A. Hobert, E.~Warmerdam, U.~Suenkel,
  K.~Brockmann, F.~Metzger, C.~Hansen, D.~Berg \emph{et~al.}, ``Gait analysis
  with wearables predicts conversion to parkinson disease,'' \emph{Annals of
  neurology}, vol.~86, no.~3, pp. 357--367, 2019.

\bibitem{curtze2015levodopa}
C.~Curtze, J.~G. Nutt, P.~Carlson-Kuhta, M.~Mancini, and F.~B. Horak,
  ``Levodopa i sa d ouble-e dged s word for b alance and g ait in p eople w ith
  p arkinson's d isease,'' \emph{Movement Disorders}, vol.~30, no.~10, pp.
  1361--1370, 2015.

\bibitem{Hodgins2008}
D.~Hodgins, ``The importance of measuring human gait.'' \emph{Medical Device
  Technology}, vol.~19, no.~5, pp. 42--44, 2008.

\bibitem{Kwapisz2011activity}
J.~R. Kwapisz, G.~M. Weiss, and S.~A. Moore, ``Activity recognition using cell
  phone accelerometers,'' \emph{ACM SigKDD Explorations Newsletter}, vol.~12,
  no.~2, pp. 74--82, 2011.

\bibitem{Bao2004activity}
L.~Bao and S.~S. Intille, ``Activity recognition from user-annotated
  acceleration data,'' in \emph{International Conference on Pervasive
  Computing}.\hskip 1em plus 0.5em minus 0.4em\relax Springer, 2004, pp. 1--17.

\bibitem{Ravi2005activity}
N.~Ravi, N.~Dandekar, P.~Mysore, and M.~L. Littman, ``Activity recognition from
  accelerometer data,'' in \emph{Aaai}, vol.~5, no. 2005, 2005, pp. 1541--1546.

\bibitem{hanlon2009real}
M.~Hanlon and R.~Anderson, ``Real-time gait event detection using wearable
  sensors,'' \emph{Gait \& posture}, vol.~30, no.~4, pp. 523--527, 2009.

\bibitem{brajdic2013}
A.~Brajdic and R.~Harle, ``Walk detection and step counting on unconstrained
  smartphones,'' in \emph{Proceedings of the 2013 ACM international joint
  conference on Pervasive and ubiquitous computing}.\hskip 1em plus 0.5em minus
  0.4em\relax ACM, 2013, pp. 225--234.

\bibitem{Williamson2000gait}
R.~Williamson and B.~J. Andrews, ``Gait event detection for fes using
  accelerometers and supervised machine learning,'' \emph{IEEE Transactions on
  Rehabilitation Engineering}, vol.~8, no.~3, pp. 312--319, 2000.

\bibitem{Sama2012}
A.~Sam{\`a}, C.~P{\'e}rez-L{\'o}pez, J.~Romagosa, D.~Rodriguez-Martin,
  A.~Catal{\`a}, J.~Cabestany, D.~Perez-Martinez, and
  A.~Rodr{\'\i}guez-Molinero, ``Dyskinesia and motor state detection in
  parkinson's disease patients with a single movement sensor,'' in \emph{2012
  Annual International Conference of the IEEE Engineering in Medicine and
  Biology Society}.\hskip 1em plus 0.5em minus 0.4em\relax IEEE, 2012, pp.
  1194--1197.

\bibitem{Tao2012}
W.~Tao, T.~Liu, R.~Zheng, and H.~Feng, ``Gait analysis using wearable
  sensors,'' \emph{Sensors}, vol.~12, no.~2, pp. 2255--2283, 2012.

\bibitem{Kavanagh2008}
J.~J. Kavanagh and H.~B. Menz, ``Accelerometry: a technique for quantifying
  movement patterns during walking,'' \emph{Gait \& posture}, vol.~28, no.~1,
  pp. 1--15, 2008.

\bibitem{del2016}
S.~Del~Din, A.~Godfrey, B.~Galna, S.~Lord, and L.~Rochester, ``Free-living gait
  characteristics in ageing and parkinson’s disease: impact of environment
  and ambulatory bout length,'' \emph{Journal of neuroengineering and
  rehabilitation}, vol.~13, no.~1, p.~46, 2016.

\bibitem{fox2009nonparametric}
E.~Fox, E.~B. Sudderth, M.~I. Jordan, and A.~S. Willsky, ``Nonparametric
  bayesian learning of switching linear dynamical systems,'' in \emph{Advances
  in Neural Information Processing Systems}, 2009, pp. 457--464.

\bibitem{lee2011activity}
Y.-S. Lee and S.-B. Cho, ``Activity recognition using hierarchical hidden
  markov models on a smartphone with 3d accelerometer,'' in \emph{International
  Conference on Hybrid Artificial Intelligence Systems}.\hskip 1em plus 0.5em
  minus 0.4em\relax Springer, 2011, pp. 460--467.

\bibitem{lukowicz2004recognizing}
P.~Lukowicz, J.~A. Ward, H.~Junker, M.~St{\"a}ger, G.~Tr{\"o}ster, A.~Atrash,
  and T.~Starner, ``Recognizing workshop activity using body worn microphones
  and accelerometers,'' in \emph{International conference on pervasive
  computing}.\hskip 1em plus 0.5em minus 0.4em\relax Springer, 2004, pp.
  18--32.

\bibitem{olguin2006human}
D.~O. Olgu{\i}n and A.~S. Pentland, ``Human activity recognition: Accuracy
  across common locations for wearable sensors,'' in \emph{Proceedings of 2006
  10th IEEE international symposium on wearable computers, Montreux,
  Switzerland}.\hskip 1em plus 0.5em minus 0.4em\relax Citeseer, 2006, pp.
  11--14.

\bibitem{Mannini2011activity}
A.~Mannini and A.~M. Sabatini, ``Accelerometry-based classification of human
  activities using markov modeling,'' \emph{Computational intelligence and
  neuroscience}, vol. 2011, p.~4, 2011.

\bibitem{Mannini2013}
A.~Mannini, S.~S. Intille, M.~Rosenberger, A.~M. Sabatini, and W.~Haskell,
  ``Activity recognition using a single accelerometer placed at the wrist or
  ankle,'' \emph{Medicine and science in sports and exercise}, vol.~45, no.~11,
  p. 2193, 2013.

\bibitem{nickel2011benchmarking}
C.~Nickel, H.~Brandt, and C.~Busch, ``Benchmarking the performance of svms and
  hmms for accelerometer-based biometric gait recognition,'' in \emph{2011 IEEE
  International Symposium on Signal Processing and Information Technology
  (ISSPIT)}.\hskip 1em plus 0.5em minus 0.4em\relax IEEE, 2011, pp. 281--286.

\bibitem{Mannini2011gait}
A.~Mannini and A.~M. Sabatini, ``A hidden markov model-based technique for gait
  segmentation using a foot-mounted gyroscope,'' in \emph{Engineering in
  Medicine and Biology Society, EMBC, 2011 Annual International Conference of
  the IEEE}.\hskip 1em plus 0.5em minus 0.4em\relax IEEE, 2011, pp. 4369--4373.

\bibitem{Wong2007}
W.~Y. Wong, M.~S. Wong, and K.~H. Lo, ``Clinical applications of sensors for
  human posture and movement analysis: a review,'' \emph{Prosthetics and
  orthotics international}, vol.~31, no.~1, pp. 62--75, 2007.

\bibitem{klucken2013unbiased}
J.~Klucken, J.~Barth, P.~Kugler, J.~Schlachetzki, T.~Henze, F.~Marxreiter,
  Z.~Kohl, R.~Steidl, J.~Hornegger, B.~Eskofier \emph{et~al.}, ``Unbiased and
  mobile gait analysis detects motor impairment in parkinson's disease,''
  \emph{PloS one}, vol.~8, no.~2, p. e56956, 2013.

\bibitem{Doherty2017}
A.~Doherty, D.~Jackson, N.~Hammerla, T.~Pl{\"o}tz, P.~Olivier, M.~H. Granat,
  T.~White, V.~T. Van~Hees, M.~I. Trenell, C.~G. Owen \emph{et~al.}, ``Large
  scale population assessment of physical activity using wrist worn
  accelerometers: the uk biobank study,'' \emph{PloS one}, vol.~12, no.~2, p.
  e0169649, 2017.

\bibitem{brognara2019assessing}
L.~Brognara, P.~Palumbo, B.~Grimm, and L.~Palmerini, ``Assessing gait in
  parkinson’s disease using wearable motion sensors: a systematic review,''
  \emph{Diseases}, vol.~7, no.~1, p.~18, 2019.

\bibitem{Randell2003}
C.~Randell, C.~Djiallis, and H.~Muller, ``Personal position measurement using
  dead reckoning,'' in \emph{Seventh IEEE International Symposium on Wearable
  Computers, 2003. Proceedings.}\hskip 1em plus 0.5em minus 0.4em\relax IEEE,
  2003, pp. 166--173.

\bibitem{Dijkstra2010detection}
B.~Dijkstra, Y.~P. Kamsma, and W.~Zijlstra, ``Detection of gait and postures
  using a miniaturized triaxial accelerometer-based system: accuracy in
  patients with mild to moderate parkinson's disease,'' \emph{Archives of
  physical medicine and rehabilitation}, vol.~91, no.~8, pp. 1272--1277, 2010.

\bibitem{Antonsson1985}
E.~K. Antonsson and R.~W. Mann, ``The frequency content of gait,''
  \emph{Journal of biomechanics}, vol.~18, no.~1, pp. 39--47, 1985.

\bibitem{Allen1977}
J.~B. Allen and L.~R. Rabiner, ``A unified approach to short-time fourier
  analysis and synthesis,'' \emph{Proceedings of the IEEE}, vol.~65, no.~11,
  pp. 1558--1564, 1977.

\bibitem{Sama2017}
A.~Sam{\`a}, C.~P{\'e}rez-L{\'o}pez, D.~Rodr{\'\i}guez-Mart{\'\i}n,
  A.~Catal{\`a}, J.~M. Moreno-Ar{\'o}stegui, J.~Cabestany, E.~de~Mingo, and
  A.~Rodr{\'\i}guez-Molinero, ``Estimating bradykinesia severity in parkinson's
  disease by analysing gait through a waist-worn sensor,'' \emph{Computers in
  biology and medicine}, vol.~84, pp. 114--123, 2017.

\bibitem{Karantonis2006}
D.~M. Karantonis, M.~R. Narayanan, M.~Mathie, N.~H. Lovell, and B.~G. Celler,
  ``Implementation of a real-time human movement classifier using a triaxial
  accelerometer for ambulatory monitoring,'' \emph{IEEE transactions on
  information technology in biomedicine}, vol.~10, no.~1, pp. 156--167, 2006.

\bibitem{mallat2008wavelet}
S.~Mallat, \emph{A Wavelet Tour of Signal Processing: The Sparse Way}.\hskip
  1em plus 0.5em minus 0.4em\relax Burlington, Massachusetts: Academic press,
  2008.

\bibitem{Figo2010}
D.~Figo, P.~C. Diniz, D.~R. Ferreira, and J.~M. Cardoso, ``Preprocessing
  techniques for context recognition from accelerometer data,'' \emph{Personal
  and Ubiquitous Computing}, vol.~14, no.~7, pp. 645--662, 2010.

\bibitem{El2018}
A.~El-Attar, A.~S. Ashour, N.~Dey, H.~Abdelkader, M.~M. Abd El-Naby, and R.~S.
  Sherratt, ``Discrete wavelet transform-based freezing of gait detection in
  parkinson’s disease,'' \emph{Journal of Experimental \& Theoretical
  Artificial Intelligence}, pp. 1--17, 2018.

\bibitem{little2019machine}
M.~A. Little, \emph{Machine Learning for Signal Processing: Data Science,
  Algorithms, and Computational Statistics}.\hskip 1em plus 0.5em minus
  0.4em\relax Oxford University Press, USA, 2019.

\bibitem{Rai2012}
A.~Rai, K.~K. Chintalapudi, V.~N. Padmanabhan, and R.~Sen, ``Zee: Zero-effort
  crowdsourcing for indoor localization,'' in \emph{Proceedings of the 18th
  annual international conference on Mobile computing and networking}.\hskip
  1em plus 0.5em minus 0.4em\relax ACM, 2012, pp. 293--304.

\bibitem{Makihara2010}
Y.~Makihara, N.~T. Trung, H.~Nagahara, R.~Sagawa, Y.~Mukaigawa, and Y.~Yagi,
  ``Phase registration of a single quasi-periodic signal using self dynamic
  time warping,'' in \emph{Asian Conference on Computer Vision}.\hskip 1em plus
  0.5em minus 0.4em\relax Springer, 2010, pp. 667--678.

\bibitem{nyan2006classification}
M.~Nyan, F.~Tay, K.~Seah, and Y.~Sitoh, ``Classification of gait patterns in
  the time--frequency domain,'' \emph{Journal of biomechanics}, vol.~39,
  no.~14, pp. 2647--2656, 2006.

\bibitem{mathie2004classification}
M.~Mathie, B.~G. Celler, N.~H. Lovell, and A.~Coster, ``Classification of basic
  daily movements using a triaxial accelerometer,'' \emph{Medical and
  Biological Engineering and Computing}, vol.~42, no.~5, pp. 679--687, 2004.

\bibitem{nickel2011using}
C.~Nickel, C.~Busch, S.~Rangarajan, and M.~M{\"o}bius, ``Using hidden markov
  models for accelerometer-based biometric gait recognition,'' in \emph{2011
  IEEE 7th International Colloquium on Signal Processing and its
  Applications}.\hskip 1em plus 0.5em minus 0.4em\relax IEEE, 2011, pp. 58--63.

\bibitem{haji2018segmentation}
N.~Haji~Ghassemi, J.~Hannink, C.~Martindale, H.~Ga{\ss}ner, M.~M{\"u}ller,
  J.~Klucken, and B.~Eskofier, ``Segmentation of gait sequences in sensor-based
  movement analysis: a comparison of methods in parkinson’s disease,''
  \emph{Sensors}, vol.~18, no.~1, p. 145, 2018.

\bibitem{Ying2007}
H.~Ying, C.~Silex, A.~Schnitzer, S.~Leonhardt, and M.~Schiek, ``Automatic step
  detection in the accelerometer signal,'' in \emph{4th International Workshop
  on Wearable and Implantable Body Sensor Networks (BSN 2007)}.\hskip 1em plus
  0.5em minus 0.4em\relax Springer, 2007, pp. 80--85.

\bibitem{Marschollek2008}
M.~Marschollek, M.~Goevercin, K.-H. Wolf, B.~Song, M.~Gietzelt, R.~Haux, and
  E.~Steinhagen-Thiessen, ``A performance comparison of accelerometry-based
  step detection algorithms on a large, non-laboratory sample of healthy and
  mobility-impaired persons,'' in \emph{2008 30th Annual International
  Conference of the IEEE Engineering in Medicine and Biology Society}.\hskip
  1em plus 0.5em minus 0.4em\relax IEEE, 2008, pp. 1319--1322.

\bibitem{Rong2007}
L.~Rong, D.~Zhiguo, Z.~Jianzhong, and L.~Ming, ``Identification of individual
  walking patterns using gait acceleration,'' in \emph{2007 1st international
  Conference on Bioinformatics and Biomedical Engineering}.\hskip 1em plus
  0.5em minus 0.4em\relax IEEE, 2007, pp. 543--546.

\bibitem{camps2018deep}
J.~Camps, A.~Sama, M.~Martin, D.~Rodriguez-Martin, C.~Perez-Lopez, J.~M.~M.
  Arostegui, J.~Cabestany, A.~Catala, S.~Alcaine, B.~Mestre \emph{et~al.},
  ``Deep learning for freezing of gait detection in parkinson’s disease
  patients in their homes using a waist-worn inertial measurement unit,''
  \emph{Knowledge-Based Systems}, vol. 139, pp. 119--131, 2018.

\bibitem{hammerla2016}
N.~Y. Hammerla, S.~Halloran, and T.~Pl{\"o}tz, ``Deep, convolutional, and
  recurrent models for human activity recognition using wearables,''
  \emph{arXiv preprint arXiv:1604.08880}, 2016.

\bibitem{Mccamley2012enhanced}
J.~McCamley, M.~Donati, E.~Grimpampi, and C.~Mazza, ``An enhanced estimate of
  initial contact and final contact instants of time using lower trunk inertial
  sensor data,'' \emph{Gait \& posture}, vol.~36, no.~2, pp. 316--318, 2012.

\bibitem{zijlstra2003assessment}
W.~Zijlstra and A.~L. Hof, ``Assessment of spatio-temporal gait parameters from
  trunk accelerations during human walking,'' \emph{Gait \& posture}, vol.~18,
  no.~2, pp. 1--10, 2003.

\bibitem{del2015validation}
S.~Del~Din, A.~Godfrey, and L.~Rochester, ``Validation of an accelerometer to
  quantify a comprehensive battery of gait characteristics in healthy older
  adults and parkinson's disease: toward clinical and at home use,'' \emph{IEEE
  journal of biomedical and health informatics}, vol.~20, no.~3, pp. 838--847,
  2015.

\bibitem{Moore2011}
S.~T. Moore, V.~Dilda, B.~Hakim, and H.~G. MacDougall, ``Validation of 24-hour
  ambulatory gait assessment in parkinson's disease with simultaneous video
  observation,'' \emph{Biomedical engineering online}, vol.~10, no.~1, p.~82,
  2011.

\bibitem{weiss2014objective}
A.~Weiss, T.~Herman, N.~Giladi, and J.~M. Hausdorff, ``Objective assessment of
  fall risk in parkinson's disease using a body-fixed sensor worn for 3 days,''
  \emph{PloS one}, vol.~9, no.~5, 2014.

\bibitem{rispens2015identification}
S.~M. Rispens, K.~S. van Schooten, M.~Pijnappels, A.~Daffertshofer, P.~J. Beek,
  and J.~H. van Dieen, ``Identification of fall risk predictors in daily life
  measurements: gait characteristics’ reliability and association with
  self-reported fall history,'' \emph{Neurorehabilitation and neural repair},
  vol.~29, no.~1, pp. 54--61, 2015.

\bibitem{perez2016assessing}
C.~P{\'e}rez-L{\'o}pez, A.~Sam{\`a}, D.~Rodr{\'\i}guez-Mart{\'\i}n,
  A.~Catal{\`a}, J.~Cabestany, J.~M. Moreno-Arostegui, E.~De~Mingo, and
  A.~Rodr{\'\i}guez-Molinero, ``Assessing motor fluctuations in parkinson’s
  disease patients based on a single inertial sensor,'' \emph{Sensors},
  vol.~16, no.~12, p. 2132, 2016.

\bibitem{bellanca2013harmonic}
J.~Bellanca, K.~Lowry, J.~VanSwearingen, J.~Brach, and M.~Redfern, ``Harmonic
  ratios: a quantification of step to step symmetry,'' \emph{Journal of
  biomechanics}, vol.~46, no.~4, pp. 828--831, 2013.

\bibitem{moe2004estimation}
R.~Moe-Nilssen and J.~L. Helbostad, ``Estimation of gait cycle characteristics
  by trunk accelerometry,'' \emph{Journal of biomechanics}, vol.~37, no.~1, pp.
  121--126, 2004.

\bibitem{Goetz2008}
C.~G. Goetz, B.~C. Tilley, S.~R. Shaftman, G.~T. Stebbins, S.~Fahn,
  P.~Martinez-Martin, W.~Poewe, C.~Sampaio, M.~B. Stern, R.~Dodel
  \emph{et~al.}, ``Movement disorder society-sponsored revision of the unified
  parkinson's disease rating scale (mds-updrs): scale presentation and
  clinimetric testing results,'' \emph{Movement disorders: official journal of
  the Movement Disorder Society}, vol.~23, no.~15, pp. 2129--2170, 2008.

\bibitem{munetz1988examine}
M.~R. Munetz and S.~Benjamin, ``How to examine patients using the abnormal
  involuntary movement scale,'' \emph{Psychiatric Services}, vol.~39, no.~11,
  pp. 1172--1177, 1988.

\bibitem{zhan2016high}
A.~Zhan, M.~A. Little, D.~A. Harris, S.~O. Abiola, E.~R. Dorsey, S.~Saria, and
  A.~Terzis, ``High frequency remote monitoring of {P}arkinson's disease via
  smartphone: Platform overview and medication response detection,''
  \emph{arXiv preprint arXiv:1601.00960}, vol. abs/1601.00960, 2016.

\bibitem{Madgwick2011}
S.~O. Madgwick, A.~J. Harrison, and R.~Vaidyanathan, ``Estimation of imu and
  marg orientation using a gradient descent algorithm,'' in \emph{2011 IEEE
  international conference on rehabilitation robotics}.\hskip 1em plus 0.5em
  minus 0.4em\relax IEEE, 2011, pp. 1--7.

\bibitem{Van2013}
V.~T. Van~Hees, L.~Gorzelniak, E.~C.~D. Leon, M.~Eder, M.~Pias, S.~Taherian,
  U.~Ekelund, F.~Renstr{\"o}m, P.~W. Franks, A.~Horsch \emph{et~al.},
  ``Separating movement and gravity components in an acceleration signal and
  implications for the assessment of human daily physical activity,''
  \emph{PloS one}, vol.~8, no.~4, p. e61691, 2013.

\bibitem{bergamini2014estimating}
E.~Bergamini, G.~Ligorio, A.~Summa, G.~Vannozzi, A.~Cappozzo, and A.~M.
  Sabatini, ``Estimating orientation using magnetic and inertial sensors and
  different sensor fusion approaches: Accuracy assessment in manual and
  locomotion tasks,'' \emph{Sensors}, vol.~14, no.~10, pp. 18\,625--18\,649,
  2014.

\bibitem{badawy2018}
R.~Badawy, Y.~P. Raykov, L.~J. Evers, B.~R. Bloem, M.~J. Faber, A.~Zhan,
  K.~Claes, and M.~A. Little, ``Automated quality control for sensor based
  symptom measurement performed outside the lab,'' \emph{Sensors}, vol.~18,
  no.~4, p. 1215, 2018.

\bibitem{kim2009ell_1}
S.-J. Kim, K.~Koh, S.~Boyd, and D.~Gorinevsky, ``$\backslash$ell\_1 trend
  filtering,'' \emph{SIAM review}, vol.~51, no.~2, pp. 339--360, 2009.

\bibitem{mckinley1998cubic}
S.~McKinley and M.~Levine, ``Cubic spline interpolation,'' \emph{College of the
  Redwoods}, vol.~45, no.~1, pp. 1049--1060, 1998.

\bibitem{Welch1967}
P.~Welch, ``The use of fast fourier transform for the estimation of power
  spectra: a method based on time averaging over short, modified
  periodograms,'' \emph{IEEE Transactions on audio and electroacoustics},
  vol.~15, no.~2, pp. 70--73, 1967.

\bibitem{allen1977short}
J.~Allen, ``Short term spectral analysis, synthesis, and modification by
  discrete fourier transform,'' \emph{IEEE Transactions on Acoustics, Speech,
  and Signal Processing}, vol.~25, no.~3, pp. 235--238, 1977.

\bibitem{Mackay1995}
D.~J. MacKay, ``Probable networks and plausible predictions—a review of
  practical bayesian methods for supervised neural networks,'' \emph{Network:
  computation in neural systems}, vol.~6, no.~3, pp. 469--505, 1995.

\bibitem{beal2003variational}
M.~J. Beal \emph{et~al.}, \emph{Variational algorithms for approximate Bayesian
  inference}.\hskip 1em plus 0.5em minus 0.4em\relax university of London
  London, 2003.

\bibitem{fox2009bayesian}
E.~B. Fox, ``Bayesian nonparametric learning of complex dynamical phenomena,''
  Ph.D. dissertation, Massachusetts Institute of Technology, 2009.

\bibitem{Kim1994}
C.-J. Kim, ``Dynamic linear models with markov-switching,'' \emph{Journal of
  Econometrics}, vol.~60, no. 1-2, pp. 1--22, 1994.

\bibitem{Beal2002}
M.~J. Beal, Z.~Ghahramani, and C.~E. Rasmussen, ``The infinite hidden {M}arkov
  model,'' in \emph{Advances in Neural Information Processing Systems 14},
  2002, pp. 577--584.

\bibitem{Teh2004}
Y.~W. Teh, M.~I. Jordan, M.~J. Beal, and D.~M. Blei, ``Sharing clusters among
  related groups: {H}ierarchical {D}irichlet processes,'' in \emph{Advances in
  Neural Information Processing Systems 17}, 2005, pp. 1385--1392.

\bibitem{Raykov2016}
Y.~P. Raykov, E.~Ozer, G.~Dasika, A.~Boukouvalas, and M.~A. Little,
  ``Predicting room occupancy with a single passive infrared {(PIR)} sensor
  through behavior extraction,'' in \emph{Proceedings of the 2016 ACM
  International Joint Conference on Pervasive and Ubiquitous Computing}, 2016,
  pp. 1016--1027.

\bibitem{barralon2006walk}
P.~Barralon, N.~Vuillerme, and N.~Noury, ``Walk detection with a kinematic
  sensor: Frequency and wavelet comparison,'' in \emph{2006 International
  Conference of the IEEE Engineering in Medicine and Biology Society}.\hskip
  1em plus 0.5em minus 0.4em\relax IEEE, 2006, pp. 1711--1714.

\bibitem{smulders2016pharmacological}
K.~Smulders, M.~L. Dale, P.~Carlson-Kuhta, J.~G. Nutt, and F.~B. Horak,
  ``Pharmacological treatment in parkinson's disease: Effects on gait,''
  \emph{Parkinsonism \& related disorders}, vol.~31, pp. 3--13, 2016.

\bibitem{hammerla2015pd}
N.~Y. Hammerla, J.~Fisher, P.~Andras, L.~Rochester, R.~Walker, and
  T.~Pl{\"o}tz, ``{PD} disease state assessment in naturalistic environments
  using deep learning,'' in \emph{Proceedings of the Twenty-Ninth AAAI
  Conference on Artificial Intelligence}, 2015, pp. 1742--1748.

\end{thebibliography}

\end{document}